\documentclass[fleqn,usenatbib]{mnras}

\usepackage{newtxtext,newtxmath}

\usepackage[T1]{fontenc}
\usepackage{ae,aecompl}

\usepackage{graphicx}	
\usepackage{amsmath}	
\usepackage{amssymb}	

\usepackage{tikz}
\usetikzlibrary{positioning}
\usetikzlibrary{arrows}

\title[IMBHs at the centre of GCs]{Dynamical modelling of globular clusters: challenges for the robust determination of IMBH candidates}

\author[Aros et al.]{
Francisco I. Aros$^{1, 2, 3}$\thanks{E-mail: francisco.aros@univie.ac.at}, Anna C. Sippel$^{3}$, Alessandra Mastrobuono-Battisti$^{4,3}$, 
\newauthor Abbas Askar$^{4}$, Paolo Bianchini$^{5}$, Glenn van de Ven$^{1,2}$
\\
$^{1}$Department of Astrophysics, University of Vienna, T\"urkenschanzstrasse 17, 1180 Vienna, Austria\\
$^{2}$European Southern Observatory (ESO), Karl-Schwarschild-Str. 2, 85748 Garching bei M\"unchen, Germany\\ 
$^{3}$Max Planck Institute for Astronomy, K\"onigstuhl 17, D-69117 Heidelberg, Germany\\
$^{4}$Lund Observatory, Department of Astronomy and Theoretical Physics, Lund University, Box 43, SE-221 00 Lund, Sweden\\
$^{5}$Observatoire Astronomique de Strasbourg, Universit\'e de Strasbourg, CNRS UMR7550, Strasbourg, France\\
}

\date{Accepted XXX. Received YYY; in original form ZZZ}

\pubyear{2019}

\begin{document}
\label{firstpage}
\pagerange{\pageref{firstpage}--\pageref{lastpage}}
\maketitle

\begin{abstract}
The presence or absence of intermediate-mass black holes (IMBHs) at the centre of Milky Way globular clusters (GCs) is still an open question. This is either due to observational restrictions or limitations in the dynamical modelling method; in this work, we explore the latter. Using a sample of high-end Monte Carlo simulations of GCs, with and without a central IMBH, we study the limitations of spherically symmetric Jeans models assuming constant velocity anisotropy and mass-to-light ratio. This dynamical method is one of the most widely used modelling approaches to identify a central IMBH in observations. 

With these models, we are able to robustly identify and recover the mass of the central IMBH in our simulation with a high-mass IMBH ($M_{\rm IMBH}/M_{\rm GC}\sim4\%$). Simultaneously, we show that it is challenging to confirm the existence of a low-mass IMBH ($M_{\rm IMBH}/M_{\rm GC}\sim0.3\%$), as both solutions with and without an IMBH are possible within our adopted error bars. For simulations without an IMBH we do not find any certain false detection of an IMBH. However, we obtain upper limits which still allow for the presence of a central IMBH. We conclude that while our modelling approach is reliable for the high-mass IMBH and does not seem to lead towards a false detection of a central IMBH, it lacks the sensitivity to robustly identify a low-mass IMBH and to definitely rule out the presence of an IMBH when it is not there. 

\end{abstract}

\begin{keywords}
globular clusters: general -- stars: kinematics and dynamics -- stars: black holes
\end{keywords}



\section{Introduction}

With masses between $10^2\,M_{\odot}$ and $10^5\,M_{\odot}$, intermediate-mass black holes (IMBHs) are still an elusive population. Ultra-luminous X-ray sources are thought to be accretion signatures of IMBHs, ESO 243-49 HLX-1 being one of the most promising candidates with a minimum mass of $500\,\text{M}_{\odot}$ \citep{farrell_2009}. Recently the gravitational wave observatories LIGO and Virgo detected a $\sim 140\,M_{\odot}$ black hole  \citep{abbott_2020a,abbott_2020b}. In the local neighbourhood a few candidates have been suggested through dynamical analysis of nearby globular clusters (GCs) \citep[see e.g. ][]{noyola_2008,van_der_marel_2010,lutzgendorf_2013a,lutzgendorf_2015}. Despite their scarce evidence, IMBHs are thought to be the missing link between stellar mass black holes (BHs, with masses of $\sim 10\,M_{\odot}$) and supermassive black holes (with masses larger than $\sim10^5\,M_{\odot}$). Furthermore it has been suggested that IMBHs could be the seeds for supermassive black holes observed at high redshifts in the early universe \citep[see e.g.][for a review]{haiman_2013}. Possible paths for the formation of IMBHs are the direct collapse of a massive star \citep{madau_rees_2001,spera_mapelli_2017} and the runaway merger of stars in dense stellar systems \citep{portegies_zwart_2004}, which happens early in the evolution of the stellar system \citep[see also][]{giersz2015}. A third path may occur later in the evolution of dense stellar systems, where an IMBH can grow from dynamical interactions \citep{giersz2015}. The latter two scenarios suggest that a dense stellar systems, such as GCs, could host a central IMBH.

GCs are bound stellar systems of $\sim10^5 - 10^6$ stars, with total masses around $5\times 10^5 \text{M}_{\odot}$.  As their name suggests, most of them have a characteristic spherical shape. GCs are compact stellar systems with half-light radii\footnote{Unless mentioned otherwise we refer to half-light radius as the the projected radius containing half of the light in the GC ($R_h$), while the half-mass radius is the 3D radius containing half of the mass in the GC ($r_{50\%}$).} of the order of a few parsecs. Their compactness and high stellar density make them bright enough to be observed, not only in our galaxy or the local group but also beyond \citep{harris_1981,brodie_strader_2006}. Given their relatively high ages, bigger than $\sim10\,\text{Gyr}$, GCs are considered the relics of the formation epoch of galaxies \citep{vandenberg_1996,carretta_2000}. The Galactic GCs half-mass relaxation times range from $\sim100\,\text{Myr}$ to $\sim10\,\text{Gyr}$ \citep[][2010 edition]{harris_1996}, making them unique systems for dynamical studies. The short relaxation times allow for mass segregation, i.e. the sorting of higher mass stars towards the cluster centre \citep{spitzer_1987}, while evolving towards a state of partial energy equipartition \citep[see][]{spitzer_1969,trenti_2013,bianchini_2016a}.

Different methods have been utilized to find IMBHs in GCs, each relying on two types of signature: accretion of gas by the IMBH or dynamical effects due the presence of the IMBH. On one hand, the accretion signatures in Galactic GCs are dim or non-existent, pointing towards possible IMBHs masses lower than $1000\,\text{M}_{\odot}$ or no IMBHs at all \citep{tremou_2018}. On the other hand, (most of) the IMBH candidates in Galactic GCs have been suggested using dynamical signatures. Stars under the direct influence of the central IMBH will follow a Keplerian potential producing a central cusp in the velocity dispersion profile of the GC \citep[][to name a few]{gebhardt_2002,noyola_2008,noyola_2010,van_der_marel_2010,lutzgendorf_2011,lutzgendorf_2012,lutzgendorf_2013a,lutzgendorf_2015,kamann_2014,kamann_2016}. 

Even with the vast literature analyzing the dynamical signatures at the centres of GCs, there is still no consensus regarding the presence or absence of IMBHs in Galactic GCs. The central cusp in velocity dispersion is limited to stars within the radius of influence\footnote{The radius of influence $r_{\text{inf}}$ is the distance from the centre of the GC where the cumulative mass of stars (and stellar remnants) is equivalent to the mass of the central IMBH, and hence depends crucially on the mass of the IMBH.} of the IMBH ($r_{\text{inf}}$), which is typically just a fraction of the core radius. Due to the small size of the radius of influence, errors in the determination of the kinematic centre or contamination by bright stars due to crowding in the centre of the GC might hamper the dynamical analysis. Using IFU data of the central region of NGC 5139 ($\omega$ Cen), \citet{noyola_2008} find evidence of a $\sim 40000\,M_{\odot}$ IMBH. For the same cluster, using a sample of proper motion from HST, \citet{van_der_marel_2010} only find an upper limit of $18000\,M_{\odot}$ for the possible IMBH. Both studies have a difference in the position of the kinematic centre, separated by $12"$ (or $\sim0.3\,\text{pc}$ at the distance of NGC 5139), which corresponds to $1\sim2$ times the $r_{\text{inf}}$, depending on the inferred IMBH mass as given above. However, using another sample of radial velocities, \citet{noyola_2010} show that the detection of the IMBH holds for the different kinematic centres. The discrepancy between both estimates could arise from either the different kind of kinematic data or modeling technique applied. Similarly in the case of NGC 6388, where  \citet{lutzgendorf_2011,lutzgendorf_2015} find evidence for an IMBH using velocity maps from integrated spectra, while \citet{lanzoni_2013} do not observe the central velocity dispersion cusp when using the radial velocities of individual stars. More recent observations from IFU with MUSE by \citet{kamann_2018} further support the presence of a central cusp in velocity dispersion. No matter which observational technique is used, the highly crowded centres of GCs add a complex observational challenge.

In addition to the observational limitations due to a small $r_{\text{inf}}$, the detection of an IMBH is also made difficult by the limitations in the dynamical models, used to actually identify an IMBH in the observational data. While usually a constant (global) mass-to-light ratio and velocity anisotropy (see Section \ref{sec:dyn}) are assumed for the dynamical models, these quantities can vary significantly in a GC. For NGC 5139 \cite{van_der_marel_2010} show how an extended dark mass due stellar remnants is also consistent with the observed velocity dispersion profile. This possibility was also recently explored by \cite{zocchi_2019} who uses a multi-mass dynamical model, based on distribution functions, to include a central cluster of stellar-mass black holes, proving that this dark extended population could also produce the central rise in velocity dispersion in NGC 5139. Using a library of \textit{N}-body simulations, \cite{baumgardt_2019} also showed that a cluster of stellar-mass black holes at the centre of NGC 5139 was favoured over a central IMBH, in particular due to their distinctive effect on the high velocity stars at the centre of the GC. A similar case was shown by \cite{mann_2019} for 47 Tuc, where a multi-mass dynamical model with a central cluster of black holes was consistent with the kinematic data, ruling out the necessity for a central IMBH suggested by \cite{kiziltan_2017}. This has been confirmed by \cite{henault-brunet_19b} with a different type of multi-mass models.

Simulations of GCs with a central IMBH provide us with a benchmark to study the observational and dynamical modelling limitations which hinder a robust detection of an IMBH via its dynamical signatures. Work in this direction has been done by \cite{de_vita_2017}. In their work, the authors explore the recovery of IMBH masses in GCs combining Monte Carlo simulations of GCs with a central IMBH \citep{askar_2017a} and mock IFU observations from \textsc{SISCO} \citep{bianchini_2015}, addressing the effects of crowding, contamination due bright stars and the cluster center. They find that, even when the actual mass profile is fully known, it is challenging to detect low-mass IMBH or rule out the IMBH solution in cases without a central IMBH. In addition, they show that when the IMBH is detected, the inferred mass is systematically underestimated. They suggest that the reason could be unquantified effects due energy equipartition and binaries.     

In this work we explore the limitations of dynamical modelling based on Jeans equations to detect a central IMBH and the feasibility of rejecting an IMBH solution when it is truly absent. For this, we will assume rather \textit{perfectly} sampled observational data from realistic simulations of GCs and analyse it with simple, but commonly used, dynamical models. We introduce a set of Monte Carlo simulations in Sections \ref{sec:mc} and \ref{sec:models} and analyze them with Jeans models\footnote{Hereafter we refer as `models' exclusively  to the dynamical models.} described in Section \ref{sec:dyn}. We focus on the limitations in the dynamical modelling itself, which assumes constant mass-to-light ratio and velocity anisotropy (see Section \ref{sec:dyn}), we apply the same modelling pipeline to the simulated GCs in Section \ref{sec:pipeline} and then analyse the result of the fittings in Section \ref{sec:results}. In Section \ref{sec:mass_constraints} we discuss the reliability of our dynamical models and we conclude with our summary in Section \ref{sec:summary}.  

\section{Methods and Model Setup}
\begin{table*}
\centering
\caption{Initial properties of simulated GCs that were used for our analysis. The first column indicates the simulation name, given by the central object at $12\,\text{Gyr}$, while the second column indicates the symbol used for refer each simulation in all figures. \textit{N} indicates the initial number of stellar systems, $f_{\text{bin}}$ provides the initial binary fraction of the cluster. All these simulations were initially \protect\citet{king1966} models, their central concentration is given by the parameter $W_{0}$. $r_{50\%}$ is the initial half-mass radius of the cluster. $r_{t}$ gives the initial tidal radius. $R_{\text{GC}}$ is the Galactocentric radius of the cluster. The final column indicates the prescription for black hole (BH) natal kick, for `Fallback' cases, BH masses and natal kicks are computed using the mass fallback prescriptions of \protect\citet{belczynski2002}. For `No Fallback' cases, BHs are given natal kicks that follow a Maxwellian distribution with $\sigma = 265\,\text{km}\text{s}^{-1}$ \protect\citep{hobbs2005}. The metallicity of all simulations was $Z=0.001$.}
\label{tab:sim-pars-0Gyr}
\begin{tabular}{lccccccccc}
\hline
\textbf{Simulation} &  \textbf{Symbol} &\textbf{N} & \textbf{$f_{\text{bin}}$} & \textbf{$W_{0}$} & \textbf{$r_{50\%}$} & \textbf{$r_{t}$} & \textbf{$R_{\text{GC}}$} & \textbf{Central Density} & \textbf{BH Natal Kicks} \\ 
 & & & [\%] & & [pc] & [pc] & [kpc] & [$M_{\odot}\text{pc}^{-3}$] & 
\\ \hline
\textit{no IMBH/BHS} & \includegraphics[scale=0.8]{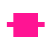} & $1.2 \times 10^{6}$ & 10 & 6 & 2.40 & 60 & 3.17 & $9.8 \times 10^{4}$ & No Fallback \\
\textit{no IMBH+BHS} & \includegraphics[scale=0.8]{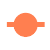} & $1.2 \times 10^{6}$ & 10 & 3 & 1.20 & 60 & 3.17 & $3.1 \times 10^{5}$ & Fallback \\
\textit{high-mass IMBH} & \includegraphics[scale=0.8]{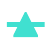} & $1.2 \times 10^{6}$ & 5 & 9 & 1.20 & 60 & 3.21 & $3.5 \times 10^{7}$ & No Fallback \\
\textit{low-mass IMBH} & \includegraphics[scale=0.8]{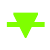} & $7.0 \times 10^{5}$ & 5 & 9 & 2.40 & 60 & 4.20 & $2.1 \times 10^{6}$ & No Fallback \\
\textit{post core-collapse} & \includegraphics[scale=0.8]{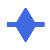} & $1.2 \times 10^{6}$ & 5 & 9 & 7.04 & 60 & 3.21 & $2.3\times 10^{5}$ & No Fallback \\ \hline
\end{tabular}
\end{table*}

We investigate the kinematic signatures of the presence of an IMBH using Monte Carlo \textit{N}-body models, evolved to $12\,\text{Gyr}$, to analyze and understand the dynamical signatures of the presence of an IMBH, as described in the following sections.

\subsection{\textsc{MOCCA} and the Monte Carlo method}
\label{sec:mc}

 The \textsc{MOCCA}-Survey Database I \citep{askar_2017a} is a collection of about 2000 simulated star clusters with different initial conditions that were evolved using the \textsc{MOCCA} code \citep[MOnte Carlo Cluster simulAtor, ][]{hypki2013,giersz2013}. The \textsc{MOCCA} code is a `kitchen sink' package that combines treatment of dynamics with prescriptions for stellar/binary evolution and other physical processes that are important in determining the evolution of a realistic star cluster. 

Dense star clusters are collisional systems and their evolution is governed by 2-body relaxation. In \textsc{MOCCA}, the treatment for relaxation is based on the orbit-averaged Monte Carlo method \citep{henon1971b,henon1971a} for following the long term evolution of spherically symmetrical star clusters. This method was subsequently improved by \citet{stod1982,stod1986} and \citet{giersz1998,giersz2001}. In this approach, relaxation is treated as a diffusive process and velocity perturbations are computed by considering an encounter between two neighboring stars. Energy and angular momentum of stars are perturbed at each timestep to mimic the effects of two-body relaxation. The Monte Carlo method combines the particle based approach of \textit{N}-body methods with a statistical treatment of relaxation. This allows for inclusion of additional physical processes that are important when simulating the evolution of a realistic star cluster. In \textsc{MOCCA}, stellar and binary evolution are implemented using the prescriptions provided by the single (\textsc{SSE}) and binary (\textsc{BSE}) codes \citep{hurley2000,hurley2002}. For computing the outcome of strong dynamical interactions involving binary-single stars and binary-binary stars, \textsc{MOCCA} uses the \textsc{FEWBODY} code \citep{fregeau2004} which was developed to carry out small-\textit{N} scattering experiments, in which case, the timestep for \textsc{FEWBODY} is set to resolve the interaction. Within one \textsc{MOCCA} timestep, many of such interactions can occur and it is also the case for binary systems interacting with an IMBH. \textsc{MOCCA} also includes a realistic treatment for the escape process in tidally limited star clusters as described by \citet{fukushige2000}. In this treatment, the escape of an object from the cluster is not instantaneous but delayed, and some potential escapers can get scattered to lower energies and become bound to the cluster again \citep{baumgardt2001}.

The main advantage of using the Monte Carlo method to simulate the dynamical evolution of a realistic star cluster is speed. \textsc{MOCCA} can compute the evolution of a million-body star cluster within a week. This advantage makes Monte Carlo codes suitable for probing the influence of the initial parameter space on the dynamical evolution of GCs. Given its underlying assumptions, the Monte Carlo method is limited to simulating spherically symmetric clusters with a timestep that is a fraction of the relaxation time. Therefore, it is well suited for following the long term evolution of a GC, but is not ideal for following the evolution on dynamical timescales. Results from \textsc{MOCCA} have been extensively compared with the results for direct \textit{N}-body simulations \citep{giersz2008,giersz2013,wang2016,madrid2017}. The evolution of global GC parameters and the number of specific objects in \textsc{MOCCA} and direct \textit{N}-body simulations are in good agreement \citep{wang2016,madrid2017}. These comparisons also serve to calibrate free parameters in the \textsc{MOCCA} code connected with the escape processes and interaction probabilities \citep{giersz2013}.  

\subsection{The Monte Carlo simulations}
\label{sec:models}

We analyze five simulated GCs with and without IMBHs, taken from the \textsc{MOCCA}-Survey Database I \citep{askar_2017a}. Their initial conditions are given in Table \ref{tab:sim-pars-0Gyr} and each is named to indicate the type of central object they contain at 12 Gyr (see also Table \ref{tab:sim-pars-12Gyr}). The \textit{no IMBH/BHS} simulation does not contain an IMBH or a significant number of BHs at 12 Gyr. The \textit{no IMBH+BHS} contains 148 stellar remnants BHs (of the order of $\sim10\,M_{\odot}$ each) at at 12 Gyr. The \textit{high-mass IMBH} cluster hosts a central IMBH of $\sim 13000\,M_{\odot}$ at 12 Gyr, while the \textit{low-mass IMBH} contains an IMBH of $\sim500\,M_{\odot}$ at 12 Gyr. The simulated cluster labeled \textit{post core-collapse} has reached core-collapse at 12 Gyr and does not contain an IMBH or a significant number of stellar mass BHs. 

All these GCs initially followed a \citet{king1966} profile and had $1.2 \times 10^{6}$ stellar systems\footnote{In this context, single and binary systems are understood as `stellar systems'. The simulated clusters start with $1.2 \times 10^{6}$ single+binary systems, rather than $1.2 \times 10^{6}$ stars.}, except for the \textit{low-mass IMBH} which initially had $7 \times 10^{5}$ stellar systems. In all cases, a metallicity of $Z=0.001$ (corresponding to $[\text{Fe}/\text{H}]\sim-1.3$) was used for the stars. The initial binary fraction for these simulated GCs is indicated in the third column in Table \ref{tab:sim-pars-0Gyr}, their initial binary properties assume a thermal eccentricity distribution, a uniform mass ratio distribution and a semi-major axis distribution which is uniform in logarithmic scale (between $2(R_{1} + R_{2})$ and $100\,\text{AU}$, where $R_{1}$ and $R_{2}$ are the zero-age main sequence stellar radii of the binary components). The simulated GCs had an initial tidal radius of $60\,\text{pc}$ and are assumed to have a circular orbit with a velocity of $220\,\text{km}/\text{s}$ around a point mass like potential for the galaxy, which total mass is equal to the enclosed mass inside the Galactocentric radius of each simulated GC (see Table \ref{tab:sim-pars-0Gyr}). 

In all simulated GCs, except the \textit{no IMBH+BHS}, BHs were given the same natal kicks as neutron stars at the moment of formation. The natal kick velocity follows a Maxwellian distribution with $\sigma = 265\, \text{km}/\text{s}$ \citep{hobbs2005}. For the \textit{no IMBH+BHS} cluster, BH masses and natal kicks were modified according to the mass fallback prescription provided by \citet{belczynski2002}. This mass fallback prescription introduces a `fall back' factor which gives the fraction of the stellar envelope that falls back on the remnant following its formation. This factor can significantly reduce natal kicks for BHs that have progenitors with zero-age main sequence masses between $20$ to $50\,M_{\odot}$. The reduced natal kicks for BHs allows the \textit{no IMBH+BHS} cluster to retain about $1300$ BHs after $50\,\text{Myr}$ of evolution. It had long been thought that BHs that are retained in GCs would efficiently eject themselves through strong dynamical interactions leaving behind at best 1 or 2 BHs up to a Hubble time \citep{sigurdsson1993,kulkarni1993}. However, recent theoretical and numerical works have shown that BH depletion might not be so efficient and GCs with moderately long relaxation times that are dynamically young could contain a sizeable number of BHs up to a Hubble time \citep{morscher2013,sippel2013,breen2013a,breen2013b,heggie2014,morscher2015,wang2016,arcasedda2018,askar2018,weatherford_2018,weatherford_2019,kremer2019}. In the same way, the presence of BHs in globular clusters has been sugested by the combination of radio and X-ray observations \citep{maccarone_2007,strader_2012,chomiuk_2013,miller-jones_2015,bahramian_2017,shishkovsky_2018,dage2018}, and kinematics \citep{giesers_2018,giesers_2019}. These observation suggest the posibility of multiple BHs in GCs. At $12\, \text{Gyr}$, the \textit{no IMBH+BHS} model has lost a significant fraction ($\sim 90\%$) of its retained BHs as the cluster evolves, but still retains about $148$ of them.

The two simulated clusters that include a central IMBH are called \textit{high-mass IMBH} and \textit{low-mass IMBH}. Both follow the formation scenarios and growth of IMBHs in GCs as seen in \textsc{MOCCA} simulations, which are described in \citet{giersz2015} and summarized in the following \citep[see also][for an analysis on all \textsc{MOCCA} simulations that include an IMBH]{arca_sedda_2019}. The \textit{high-mass IMBH} cluster had initially a central density of $3.5\times10^{7}\,M_{\odot}\text{pc}^{-3}$. Typically, for simulations with such high central densities, runaway mergers of main sequence stars in the first $50\,\text{Myr}$ lead to the formation of massive main sequence stars which can then form an IMBH seed either through a merger or collision with a stellar mass BH or through direct collapse \citep[see e.g.][]{portegies_zwart_2004,spera_mapelli_2017}. This formation scenario occurs early in the evolution of the GC and is described as the `FAST' scenario in \citet{giersz2015}. On the other hand, in the model \textit{low-mass IMBH} model, the IMBH forms after more than $9\,\text{Gyr}$ of cluster evolution via the `SLOW' scenario described in \citet{giersz2015}. In this scenario, the IMBH forms from the growth of a stellar mass BH by mergers and collisions during the core collapse stage of cluster evolution. The IMBH formed via the `SLOW' scenario have masses in the range of $10^{2} - 10^{3}\,M_{\odot}$ at $12\,\text{Gyr}$. Both simulations with a central IMBH do not have any stellar BHs within $r_{50\%}$, because the IMBH efficiently ejects or merges with stellar mass BHs in the cluster \citep{leigh2014,giersz2015}. 

The channel of formation also has an impact on the interaction between the IMBH and the surrounding stars. IMBHs formed early on through the `FAST' scenario produce a more clear central rise in velocity dispersion, while an IMBH formed via the `SLOW' scenario could lack such clear features at $12\,\text{Gyr}$, as it forms later on during the evolution of the GC \citep{giersz2015}. In principle, in MOCCA simulations, a low-mass IMBH can wander around the centre of the cluster, which in turn can hamper the formation of the velocity dispersion cusp. As the IMBH mass grows its movement around the center decreases, and it should stay fixed for IMBHs with $M_{\bullet}>1000\,M_{\odot}$. In MOCCA simulations, IMBHs with $M_{\bullet} > 1000 \sim 2000\,M_{\odot}$ should produce a clear central rise in the velocity dispersion and surface brightness profiles \citep{giersz2015}.

At $12\,\text{Gyr}$ the IMBH in the \textit{low-mass IMBH} simulation is almost the innermost object, however, its displacement with respect to the cluster centre is small ($2\times10^{-4}\,\text{pc}$ or $\sim 10\,\text{mas}$ at $5\,\text{kpc}$) and it should not have an effect in the dynamical models. However, as pointed out by \cite{de_vita_2018}, through direct \textit{N}-body simulations, large displacements of an IMBH with respect the cluster centre will require tailored data-modelling comparisons and dynamical models under the assumption of spherical symmetry (as the one used in this work and described below in Section \ref{sec:dyn}) might introduce a bias on the estimated masses of the IMBH.

The \textit{post core-collapse} simulation starts out as tidally-filling, with a half-mass radius of $\sim7\,\text{pc}$. The cluster undergoes stronger mass loss due to tidal stripping which decreases the number of stars and shortens its relaxation time. Therefore, the cluster is dynamically older and has evolved to a post core-collapse phase at $12\,\text{Gyr}$. 

For all the five simulated GCs, we extracted the 12 Gyr \textsc{MOCCA} snapshot which contains the radial position, radial velocity, tangential velocity and stellar parameters of each star. The details of how the snapshot was used for our dynamical modelling is provided in subsequent sections. In Table \ref{tab:sim-pars-12Gyr}, we provide the $12\,\text{Gyr}$ properties of each of the five simulated clusters. We have included in this table, the total mass of the cluster ($M_{\text{tot}}$), its half-mass ($r_{50\%}$) and half-light radii ($R_h$), total luminosity ($L_{\text{tot}}$), binary fraction ($f_{\text{bin}}$), mass-to-light ratio within the half-mass radius ($\Upsilon_{50\%}$), the inner ($\beta_{50\%}$) and outer velocity anisotropy ($\beta_{\text{out}}$, see Equation \ref{eq:ani}), the mass of the central IMBH ($M_{\bullet}$) and the total mass in stellar BHs ($M_{\text{bh}}$) within the half-mass radius.

\begin{table*}
\centering
\caption{Summary of the properties of the simulated GCs at $12\,\text{Gyr}$. These values were measured directly from the simulations. The first column indicates the simulation name, given by the central object at $12\,\text{Gyr}$, while the second column indicates the symbol used for refer each simulation in all figures. The number of stellar systems (N) includes single and binaries stars. $M_\text{tot}$ is the total mass of the cluster and $r_{50\%}$ is the half-mass radius, while $L_\text{tot}$ is the total cluster luminosity and $R_h$ is the projected half-light radius. The binary fraction ($f_{\text{bin}}$) represent the global fraction including all stellar systems in the simulation. The half-mass mass-to-light ratio ($\Upsilon_{50\%}$) and the half-mass velocity anisotropy ($\beta_{50\%}$) were measured including all stellar systems within the half-mass radius ($r_{50\%}$), while the outer velocity anisotropy ($\beta_{\text{out}}$) includes all stars with radii larger than $r_{50\%}$. $M_{\bullet}$ is the mass of the central IMBH, while $M_{\text{bh}}$ is the total mass of stellar black holes within $r_{50\%}$.}
\label{tab:sim-pars-12Gyr}
\begin{tabular}{|l|c|c|c|c|c|c|c|c|c|c|c|c|}
\hline
\textbf{Simulation} & \textbf{Symbol} & \textbf{N} & $M_{\text{tot}}$ & $r_{50\%}$ & $L_{\text{tot}}$ & $R_{h}$ &  $f_{\text{bin}}$ & $\Upsilon_{50\%}$ & $\beta_{50\%}$ & $\beta_{\text{out}}$ & $M_{\bullet}$ & $M_{\text{bh}}$  \\
&  &  & $\left[\times 10^5\, M_{\odot}\right]$ & $\left[ \text{pc}\right]$ & $\left[\times10^5\,L_{\odot}\right]$ & $\left[ \text{pc}\right]$ &  $\left[\%\right]$ & $\left[M_{\odot}/L_{\odot}\right]$ & & & $\left[M_{\odot}\right]$ & $\left[M_{\odot}\right]$\\  
\hline
\textit{no IMBH/BHS} &\includegraphics[scale=0.8]{FIGURES/symbols/snp_01.pdf} & 1048918 & 3.56 & 5.29 & 1.99 & 2.50 & 6.8 & 1.38 & 0.03 & 0.12 & 0.0 & 39.98\\
\textit{no IMBH+BHS} &\includegraphics[scale=0.8]{FIGURES/symbols/snp_02.pdf} & 971004 & 3.29 & 4.99 & 1.81 & 2.84 & 5.7 & 1.39 & 0.11 & 0.37 & 0.0 & 1437.61 \\
\textit{high-mass IMBH} &\includegraphics[scale=0.8]{FIGURES/symbols/snp_03.pdf} & 942585 & 3.07 & 5.50 & 1.81 & 2.63 & 2.0 & 1.26 & 0.10 & 0.30 & 12883.4 & 0.0\\
\textit{low-mass IMBH} &\includegraphics[scale=0.8]{FIGURES/symbols/snp_04.pdf} & 496159 & 1.70 & 6.13 & 0.95 & 2.02 & 3.0 & 1.40 & 0.04 & 0.08 & 519.3 & 0.0 \\
\textit{post core-collapse} &\includegraphics[scale=0.8]{FIGURES/symbols/snp_05.pdf} & 388631 & 1.42 & 5.14 & 0.83 & 1.91 & 3.7 & 1.24 & 0.00 & -0.03  & 0.0 & 15.60\\
\hline 
\end{tabular}
\end{table*}

\subsection{Dynamical modelling}
\label{sec:dyn}
We build dynamical models to characterize the 3D mass profile of the simulated GCs. Our models are built by solving the Jeans equations \citep{jeans_1922}, which allows us to characterize the internal dynamical state of a stellar system via the velocity moments of its distribution function (DF) $f(\mathbf{x},\mathbf{v})$. The following description of the Jeans equations is based on Chapter 4 of \cite{binney_2008} and Section 2 of \cite{van_der_marel_2010}.

The dynamical state of a collisionless system is fully determined by the Collisioneless Boltzmann Equation:

\begin{equation}
    \frac{\partial f}{\partial t} + \sum_{i=1}^3{\left( v_i \frac{\partial f}{\partial x_i} - \frac{\partial \Phi}{\partial x_i}\frac{\partial f}{\partial v_i} \right)} = 0\,, 
    \label{eq:boltzmann}
\end{equation}
which represents the conservation of the probability of finding a star within the phase-space of position $\mathbf{x}$ and velocity $\mathbf{v}$ given the DF $f(\mathbf{x},\mathbf{v})$ and the potential $\Phi$. However, solving and relating Equation \ref{eq:boltzmann} to observable quantities is not trivial. A simpler approach is to integrate Equation \ref{eq:boltzmann} over the velocity space assuming the system is in equilibrium ($\partial f/\partial t = 0$). This provides a set of equations, known as Jeans equations, depending only on the velocity moments, rather than on the more complex DF. The zeroth velocity moment will correspond to the probability of finding a star at a certain position $\nu(\mathbf{x})$. This is not a direct observable and it has to be evaluated using either the number density $n(\mathbf{x}) = N_{\text{tot}}\nu(\mathbf{x})$ or the luminosity density $j(\mathbf{x})=L_{\text{tot}}\nu(\mathbf{x})$ as proxies (where $N_{\text{tot}}$ and $L_{\text{tot}}$ are the total number of stars and total luminosity). Here we use the latter as proxy of the zeroth velocity moment and express all the equations below in terms of $j(\mathbf{x})$ rather than $\nu(\mathbf{x})$. The first velocity moment is the mean velocity $\langle \text{v} \rangle$, while the second velocity moment $\langle \text{v}^2 \rangle = \sigma^2 + {\langle \text{v} \rangle}^2$ includes the effects of the velocity dispersion $\sigma$ and the mean velocity $\langle \text{v} \rangle$. 

We build spherically symmetric dynamical models by assuming a DF that depends only on the Hamiltonian $H(\mathbf{x},\mathbf{v})$ and the total angular momentum $\text{L}$. For these models, the first velocity moments are ${\langle \text{v}_{r} \rangle} = 0$, $\langle\text{v}_{\varphi}\rangle = 0$ and $\langle\text{v}_{\theta}\rangle = 0$, while for the second velocity moments ${\langle \text{v}_{\varphi}^2 \rangle}={\langle \text{v}_{\theta}^2 \rangle}$ holds. This allows to define a tangential component as ${\langle \text{v}_{t}^2 \rangle} = {\langle \text{v}_{\theta}^2 \rangle}+{\langle \text{v}_{\phi}^2 \rangle}$ and have an expression for the Jeans equation, which depends only of two unknowns variables ${\langle \text{v}_{r}^2 \rangle}$ and  ${\langle \text{v}_{t}^2 \rangle}$:

\begin{equation}
\frac{d}{dr}\left({j(r)\langle \text{v}_r^2 \rangle}\right) + j(r)\left( \frac{d\Phi}{dr} + \frac{2{\langle \text{v}_{r}^2 \rangle}-{\langle \text{v}_{t}^2 \rangle}}{r}\right)=0\,.
\label{eq:jeans_full}
\end{equation}

The dependency of the second velocity moments ${\langle \text{v}_{r}^2 \rangle}$ and  ${\langle \text{v}_{t}^2 \rangle}$ is usually described by the velocity anisotropy $\beta$ as:
\begin{equation}
\beta = 1 - \frac{{\langle \text{v}_{t}^2 \rangle}}{2{\langle \text{v}_{r}^2 \rangle}}\,,
\label{eq:ani}
\end{equation} 
\citep[see][]{binney_2008} which could take any functional form and allows us to rewrite Equation \ref{eq:jeans_full} as follows:
\begin{equation}
\frac{d}{dr}\left({j(r)\langle \text{v}_r^2 \rangle}\right) + \left({j(r)\langle \text{v}_r^2 \rangle}\right)\left(\frac{2\beta}{r}\right) = -  \frac{d\Phi}{dr} \,.
\label{eq:jeans_vr}
\end{equation}
In our case, we assume a constant velocity anisotropy through the stellar system, under this condition the second velocity moment ${\langle \text{v}_{r}^2 \rangle}$ is: 
\begin{equation}
\langle \text{v}^2_r \rangle (r) = \frac{1}{j(r) r^{2\beta}}\int_{r}^{\infty}{dr'j(r')r'^{(-2\beta)}\frac{\partial\Phi}{\partial r'}(r')}\,.
\label{eq:vrms_r}
\end{equation} 

The expression for $\langle \text{v}^2_r \rangle$ is embedded into the coordinate system centred in the stellar system, but as external observers we usually do not have the full 6-dimensional information (i.e. the three position and three velocities). At most we have available the individual position of each star projected in the sky $(x',y')$, the line-of-sight velocity $(\text{v}_{\text{LOS}})$, the radial $(\text{v}_{\text{PMR}})$ proper motion and the tangential $(\text{v}_{\text{PMT}})$ proper motion. These are shown in Figure \ref{fig:sky_coord}.

To relate  $\langle\text{v}^2_r\rangle$  with the observations we integrate it along the line-of-sight to get a weighted average for the second velocity moments:

\begin{figure}
    \centering
    \includegraphics[width=0.8\linewidth]{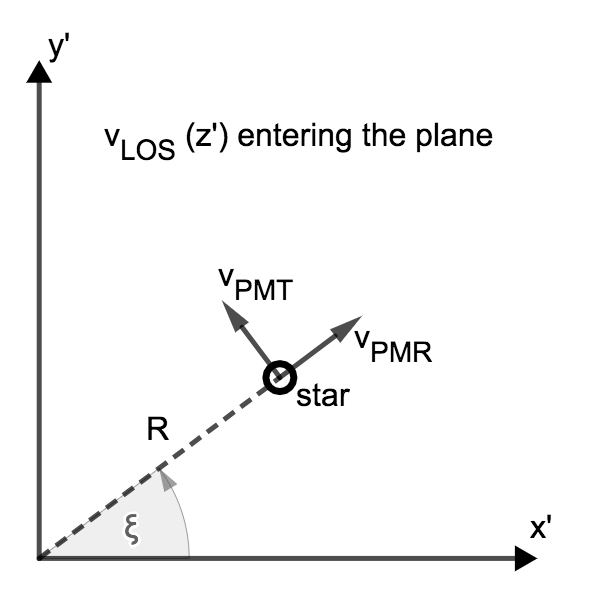}
    \caption{Sky coordinates for the projected velocity components. The star is located at a projected distance $R$ from the cluster centre in the plane of the sky ($x'y'$ plane). The line-of-sight velocity ($\text{v}_{\text{LOS}}$) is perpendicular to the plane of the sky, while the radial proper motion ($\text{v}_{\text{PMR}}$) follows the direction of the radial vector defined by $R$, and the tangential proper motion ($\text{v}_{\text{PMT}}$) follows the direction of the $\xi$ angle between $R$ and $x'$.}
    \label{fig:sky_coord}
\end{figure}

\begin{align}
\langle\text{v}^2_{\text{LOS}}\rangle(R) & = \frac{1}{I(R)}\displaystyle\int_{R}^{\infty} \frac{j(r) dr}{\sqrt{r^2-R^2}}\left(1-\beta\left(\frac{r}{R}\right)^2\right)\langle\text{v}^2_{r}\rangle\,,\label{eq:v2los} \\
\langle\text{v}^2_{\text{PMR}}\rangle(R) & = \frac{1}{I(R)}\displaystyle\int_{R}^{\infty} \frac{j(r) dr}{\sqrt{r^2-R^2}}\left(1-\beta+\beta\left(\frac{r}{R}\right)^2\right)\langle\text{v}^2_{r}\rangle\,,\label{eq:v2pmr} \\
\langle\text{v}^2_{\text{PMT}}\rangle(R) & = \frac{1}{I(R)}\displaystyle\int_{R}^{\infty} \frac{j(r) dr}{\sqrt{r^2-R^2}}\left(1-\beta\right)\langle\text{v}^2_{r}\rangle\,,\label{eq:v2pmt}
\end{align}
where $R = \sqrt{x'^2+y'^2}$ is the radial distance projected in the sky from the centre of the GC to the star and $I(R)$ is the surface brightness of the GC. 
We model the surface brightness in a similar way as \citep{van_der_marel_2010}, using the following function:
\begin{equation}
I(R) = I_{0}\times(R/a_0)^{-s_0}\times(1+(R/a_0)^{\alpha_1})^{-s_1/\alpha_1}\times(1+(R/a_1)^{\alpha_2})^{-s_2/\alpha_2}\,,
\label{eq:surf_lumn}
\end{equation}
where, $I_{0}$ is a scaling factor, $a_0$ and $a_1$ are the inner and outer scale radii, $s_0$ gives the slope of a possible central cusp, while $s_1$, $s_2$ and $\alpha_1$, $\alpha_2$ control the mid and outer slopes. This parametric form allows us to to explore a broad range of surface luminosity profiles and easily perform a deprojection to get the luminosity density:
\begin{equation}
j(r) = \frac{-1}{\pi}\displaystyle\int_r^{\infty}{\frac{dR}{\sqrt{R^2-r^2}}\frac{dI}{dR}}\,.\label{eq:j_r} 
\end{equation}

To determine the internal mass density profile, we assume a constant mass-to-light ratio $\Upsilon_0$ and define the stellar mass density profile as $\rho_{\star}(r) = \Upsilon_0 j(r)$. This simplification is commonly adopted. The total mass of the GC contained within the radius $r$ is then $M(r) = M_{\bullet} + M_{\star}(r)$, where $M_{\bullet}$ is the mass of the possible central black hole and $M_{\star}(r)$ is the stellar mass given by:
\begin{equation}
M_{\star}(r) = 4\pi\displaystyle\int_{0}^{r}\rho_{\star}(r')r'^2 dr'\,.\label{eq:M_stars}
\end{equation}
We express the derivative of the potential $\Phi$ as:
\begin{equation}
\frac{d\Phi}{dr} = \frac{GM_{\bullet}}{r^2} + \frac{GM_{\star}(r)}{r^2}\,,\label{eq:dPhi}
\end{equation}
where the potential will have a Keplerian component given by the central black hole mass ($M_{\bullet}$) and an extended component given by the mass distribution of stars ($M_{\star}$).

\section{Analysis and Results}

\subsection{Pipeline}
\label{sec:pipeline}

For all the different data sets mentioned in Section \ref{sec:mc} and Table \ref{tab:sim-pars-12Gyr}, we have applied the following blind approach, also summarized in Figure \ref{fig:pipeline}:
\begin{itemize}

\tikzstyle{squarebox_large} = [rectangle, minimum width=4.5cm, minimum height=1cm,text centered,text width=4.5cm, draw=black]

\tikzstyle{squarebox_small} = [rectangle, minimum width=2.5cm, minimum height=1cm,text centered,text width=2.5cm, draw=black]

\tikzstyle{roundbox} = [rectangle, rounded corners, minimum width=3.0cm, minimum height=2cm, text centered, text width=3cm, draw=black]

\tikzstyle{arrow} = [thick,->,>=stealth]

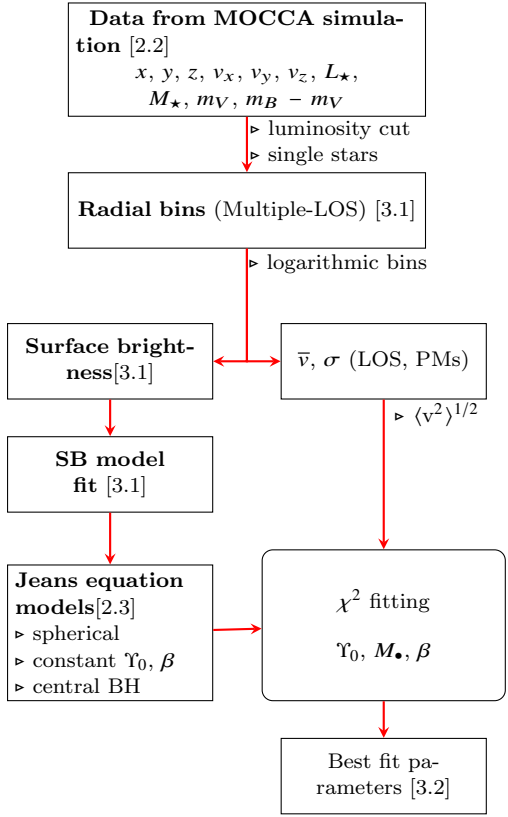
\begin{figure}
    \centering

\begin{tikzpicture}[node distance=2.0cm]

\node (data) [squarebox_large] {\textbf{Data from MOCCA simulation} [\ref{sec:models}]
                                \newline $x$, $y$, $z$, $v_x$, $v_y$, $v_z$, $L_{\star}$, $M_{\star}$, $m_{V}$, $m_{B}-m_{V}$};
\node (binning) [squarebox_large, below of=data] {\textbf{Radial bins} (Multiple-LOS) [\ref{sec:pipeline}]};

\node (lum_prof) [squarebox_small, below of=binning, xshift=-1.8cm] {\textbf{Surface brightness}[\ref{sec:pipeline}]};
\node (sb_model) [squarebox_small, below of=lum_prof,yshift=+0.5cm] {\textbf{SB model fit} [\ref{sec:pipeline}]};
\node (jeans_eq) [squarebox_small, below of=sb_model,yshift=-0.1cm,align=left] {\textbf{Jeans equation models}[\ref{sec:dyn}] \\ $\triangleright$ spherical
                                \\ $\triangleright$ constant $\Upsilon_0$, $\beta$
                                \\ $\triangleright$ central BH};

\node (kin_data) [squarebox_small, below of=binning, xshift=+1.8cm] {$\overline{v}$, $\sigma$ (LOS, PMs)};
\node (chi2_fit) [roundbox, below of=kin_data, yshift=-1.5cm,, align=center] {$\chi^2$ fitting \\ \vspace{0.3cm}  $\Upsilon_0$, $M_{\bullet}$, $\beta$ };

\node (best_par) [squarebox_small, below of=chi2_fit] {Best fit parameters [\ref{sec:results}]};

\draw [arrow,red] (data) --  (binning);
\draw [arrow,red] (binning) |- (lum_prof);
\draw [arrow,red] (binning) |- (kin_data);

\draw [arrow,red] (lum_prof) -- (sb_model);
\draw [arrow,red] (sb_model) -- (jeans_eq);
\draw [arrow,red] (kin_data) -- (chi2_fit);
\draw [arrow,red] (jeans_eq) -- (chi2_fit);
\draw [arrow,red] (chi2_fit) -- (best_par);

\node [align=left] (text_1) at (1.1,-1.1) {$\triangleright$ luminosity cut \\ $\triangleright$ single stars };
\node [align=left] (text_2) at (1.2,-2.7) {$\triangleright$ logarithmic bins};
\node [align=left] (text_3) at (2.5,-4.7) {$\triangleright$ $\langle\text{v}^2\rangle^{1/2}$};

\end{tikzpicture}

   \caption{Pipeline for the dynamical analysis of the simulated GCs as described in Section \protect\ref{sec:pipeline}. We start by extracting the required data from to simulated GCs, projected in the sky, from which we generate surface brightness and kinematic radial profiles. The surface brightness profile is used as an input for the dynamical models, which in turn are fitted to the kinematic profiles.}
    \label{fig:pipeline}
\end{figure}

\item[(1)]{
For each GC we select a subsample of stars as our kinematic tracers. The selection, which is the same for each of the GCs, impose a luminosity cut and the exclusion of all binary systems. 

We selected all stars brighter than one magnitude below the main sequence turn-off as kinematic tracers, which is equivalent to select stars brighter than $m_V = 18.5\,\text{mag}$ at a distance of $D=5\,\text{kpc}$ (without extinction). As shown in Figure \ref{fig:LUM_CUT} for the \textit{no IMBH/BHS} simulation, this selection excludes most of the stellar main-sequence along with the white dwarf sequence and fainter remnants (neutron stars and stellar black holes). Our magnitude cut resembles the fainter limit adopted by \citet{watkins_15} for HST proper motions of galactic GCs, however, astrometric catalogs can achieve even fainter magnitudes at the central \citep[see][for HST proper motions]{anderson_10,libralato_2018} and outer regions of GCs \citep[][for HST and Gaia proper motions respectively]{heyl_17,bianchini_2020}. On the other hand, while state-of-the-art line-of-sight observations are pushing towards fainter magnitudes, below the main sequence turn-off \cite[e.g. MUSE][]{giesers_2019}, their observational errors are still large compared to the typical velocity dispersion of GCs. The magnitude cut is agreement with such limitations and allows us to compare line-of-sight velocities and proper motions of our selected kinematic tracers. We have included in Figure \ref{fig:msto_lum_cut}, in the appendix, the color-magnitude diagrams for all five simulated GCs.

\begin{figure}
\begin{center}
\includegraphics[width=0.9\linewidth]{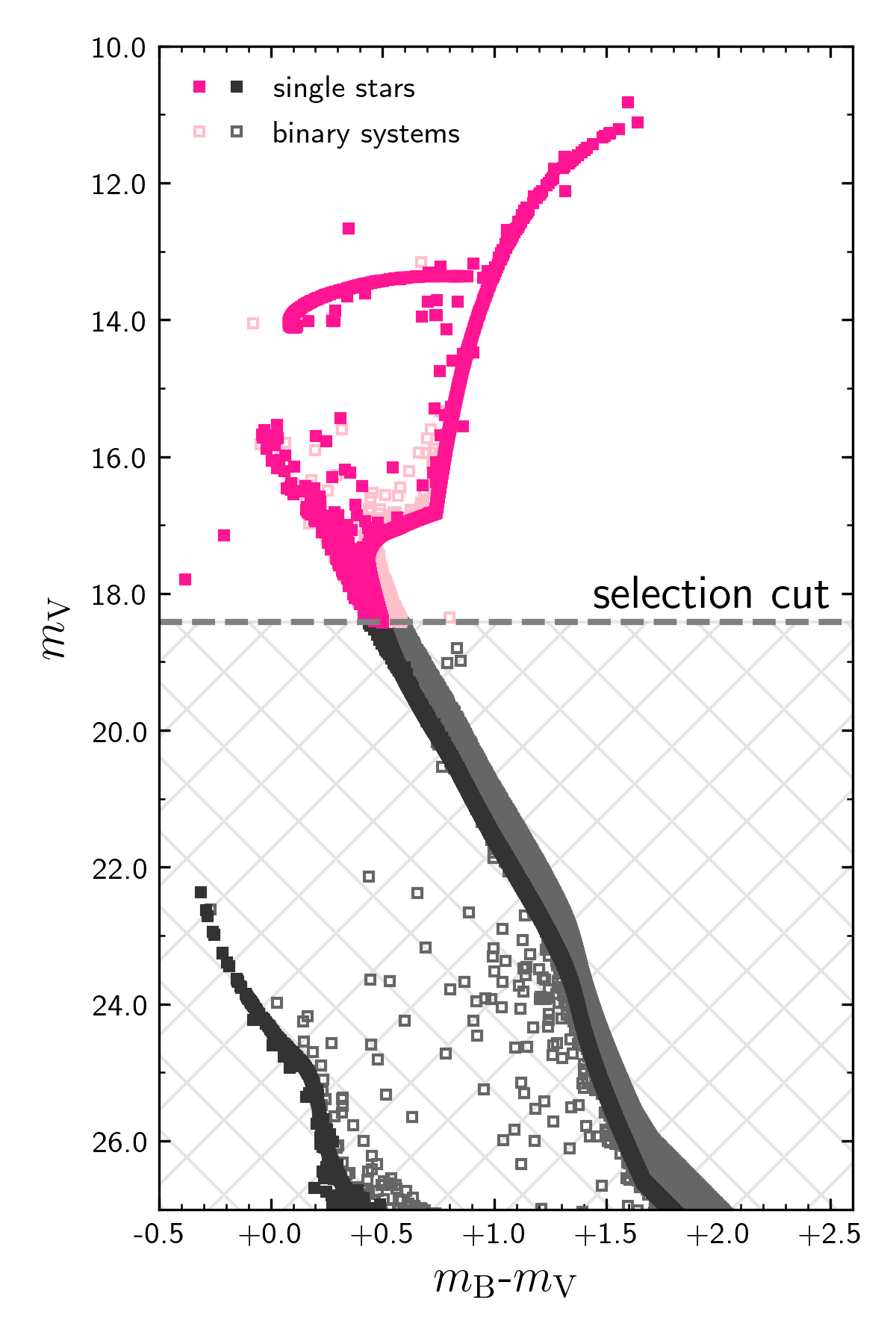}
\caption{Color-Magnitude diagram for the \textit{no IMBH/BHS} simulation. Single stars are represented by filled symbols, while binary systems are represented by open symbols. We impose a luminosity cut by selecting all stars brighter than one magnitude below the main sequence turn-off (or an apparent magnitude of $m_V\sim18.5\,\text{mag}$ at a distance of $D=5\,\text{kpc}$, without extinction). This limit is consistent with current observations of line-of-sight velocities and it excludes the most main-sequence stars, the white dwarf sequence, neutron stars and stellar black holes in the cluster.}
\label{fig:LUM_CUT}
\end{center}
\end{figure}

Within the selected sample of stellar systems in each simulation, a fraction of them will correspond to binary systems (as shown by the open squares in Figure \ref{fig:LUM_CUT}). Binary stars will have different effects in the measured velocity dispersion depending on the type of kinematic sample. For line-of-sight velocities the observed radial velocity will be dominated by the orbital velocity of the brightest component rather than their centre of mass velocity, this additional velocity will increase the measured velocity dispersion. Panel (a) of Figure \ref{fig:BIN_CENTER} shows the effect of the binary systems (open squares) in the line-of-sight velocity dispersion compared to a sample that exclude all binaries (filled squares). The individual velocities of each binary component were projected using the \textsc{COCOA}\footnote{\url{https://github.com/abs2k12/COCOA}} code \citep{askar_2018}, then we used the luminosity weighted velocity for each binary system. The bias produced by the orbital velocities of each binary system increases towards the centre of the cluster where binaries become harder.

Panel (b) in Figure \ref{fig:BIN_CENTER} shows the effects in the line-of-sight velocity dispersion for different populations of binary systems, the short period binaries ($P<30\,\text{days}$) dominates the rise in velocity dispersion observed in panel (a), while the long period binaries ($P\geq1\,\text{year}$), which do not have a large amplitude in their orbital velocity, have a shallower effect. On the other hand, proper motion velocities will not be significantly affected by the orbital motion of the binary system, as the observations will follow the velocity of the centre of mass.

However, as binary system are more massive than single stars, they will have a systematically lower velocity dispersion than single stars because of partial energy equipartition effects \citep[see][for a discussion]{bianchini_2016b}. As we expect a larger fraction of binaries towards the centre due mass segregation, the binary systems will bias the measured velocity dispersion to a lower value (see Figure \ref{fig:delta_vdisp}). This will equally affect line-of-sight velocities and proper motions.

Identifying all binaries and excluding them is not usually possible and a few contaminants might remain in real observational samples, even more given our luminosity cut. However, efforts in the direction to identify binary systems in GCs have been done \citep[see for example][]{milone_2012,giesers_2019,belokurov_2020}. The different effects of binaries on the measured velocity dispersion are highly non-trivial and might play against a robust determination of the presence of an IMBH. In this work we explicitly focus on the limitation introduced by the dynamical modelling in the IMBH mass assessment, and we leave for a follow up contribution the detailed study of the complex interplay between presence of binaries and observational biases. Furthermore, the sample without binaries is, within errors, still consistent with the sample that only includes long period binaries, which are more likely to be misidentified with line-of-sight multi epoch observations. For this reason we have excluded all binary systems from our kinematic sample in the current analysis.

\begin{figure}
    \centering
    \includegraphics[width=0.8\linewidth]{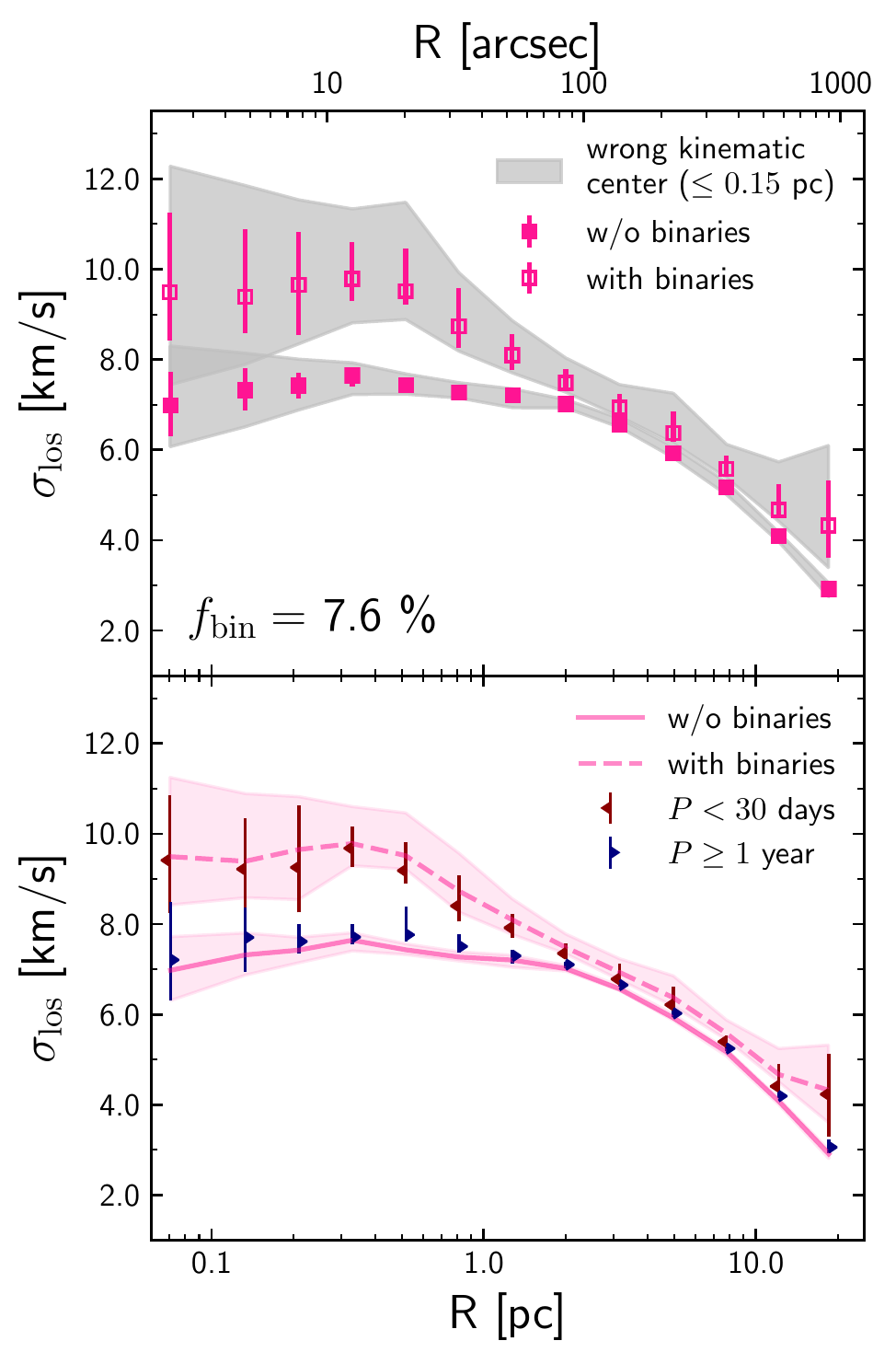}
    \caption{Line-of-sight velocity dispersion for the \textit{no IMBH/BHS} simulation. The simulated GCs have a non-negligible fraction of binary systems which can increase the observed line-of-sight velocity dispersion, as their measured radial velocity will be dominated by their orbital velocity rather than their centre of mass velocity. The binary systems become harder as their sink towards the centre of the GC.  Their intrinsic orbital velocity get larger and its effect in the observed velocity dispersion becomes more significant. Panel (a) shows the measured velocity dispersion for the selected stellar systems (as in Figure \protect\ref{fig:LUM_CUT}). The sample with binary systems (open squares) has a systematically larger velocity dispersion than the sample which only considers single stellar systems (solid squares), this difference increases towards the center where it becomes $\sim 2\,\text{km/s}$. The gray shaded areas show the effect on the velocity dispersion caused by an error in the kinematic centre up to $\text{R}=0.15\,\text{pc}$ (or $\sim 6\,\text{arcsec}$ at a distance of $5\,\text{kpc}$), this is equivalent to $20\%$ of the core radius of the GC. Not all binary system have the same influence in the measured velocity dispersion, this is shown in panel (b). Short period binaries (with $P<30\,\text{days}$, left-side triangles) dominate the increase in velocity dispersion, while binaries with longer periods ($P\geq 1\,\text{year}$, right-side triangles) do not add a significant bias into the velocity dispersion, being similar to the case without binaries. The binary fraction in the selected sample is $f_{\text{bin}} = 7.8\%$ while the fraction of binary stellar system that fall into the short period binaries is only $f_{\text{bin}} = 2\%$. The shaded areas in panel (b) represent the error bars for the samples without binaries and with all binaries.}
    \label{fig:BIN_CENTER}
\end{figure}

}

\item[(2)]{Crowding and the determination of the kinematic centre are two observational effects that have played against the robust determination of IMBHs in GCs \citep{noyola_2008,van_der_marel_2010,lutzgendorf_2013a,lanzoni_2013,de_vita_2017}. In the case of the former we assume that we can resolve all stars in the selected sample, while for the centre we use the same centre for the luminosity and kinematics}. The grey shaded area in panel (a) of Figure \ref{fig:BIN_CENTER} shows the effects in the measured velocity dispersion due an error in the kinematic centre determination up to $0.15\,\text{pc}$, approximately $20\%$ of the GC core radius \cite[see][]{de_vita_2017}. In comparison the determination of the centre in NGC 5139 is $\sim10\%$ of its core radius \citep{noyola_2010}.

\item[(3)]{With the selected sample we generate radial profiles using the projected data in the $(x,y)$ plane. The profiles follow fixed logarithmic radial bins, which allow us to have information in the central region without requiring an excessive number of bins. Using a fixed binning, and therefore having a varying number of tracers per bin, could potentially lead to low statistics, especially in the central bins. We manage the effect of low statistics by observing the GC from different line-of-sights. As the simulations have spherical symmetry, this approach allows us to have a distribution of values for each bin without altering the intrinsic radial profiles. We sampled $1000$ different line-of-sights uniformly distributed in a spherical shell, then for each bin we adopt the median to build the radial profiles and the $16^{\text{th}}$, and $84^{\text{th}}$ percentiles as an error bar (as the distribution is not necessarily symmetric). Our approach is a simplified version of the projection method described by \cite{mashchenko_2005}, where the probability of each particle to be found in a given bin is calculated as if it were observed from all line-of-sights.

Figure \ref{fig:PRJ_PRF} shows the luminosity surface density $\text{L}(\text{R})$, mass surface density $\Sigma(\text{R})$, the mean line-of-sight velocity $\overline{\text{v}}(\text{R})$ and line-of-sight velocity dispersion $\sigma(\text{R})$ profiles for the \textit{no IMBH/BHS} simulation (pink squares). As a comparison we also include the profiles when all single stars are considered (black diamonds). No major differences are observed regarding the luminosity surface density, as both samples are dominated by the same bright stars (panel (a) in Figure \ref{fig:PRJ_PRF}). The mass surface density of the selected sample is significantly lower than the full sample of single stars, as our selected sample only adds up to the $4.2\%$ of the total mass of the simulated \textit{no IMBH/BHS} cluster. The velocity dispersion is lower in our selected sample within $R_h$, which is an expected effect of energy equipartition \cite[see e.g. ][]{trenti_2013,bianchini_2016a}. It is important to be aware of these differences, as our tracers do not provide the full information about the mass profile of the cluster.

\begin{figure}
    \centering
    \includegraphics{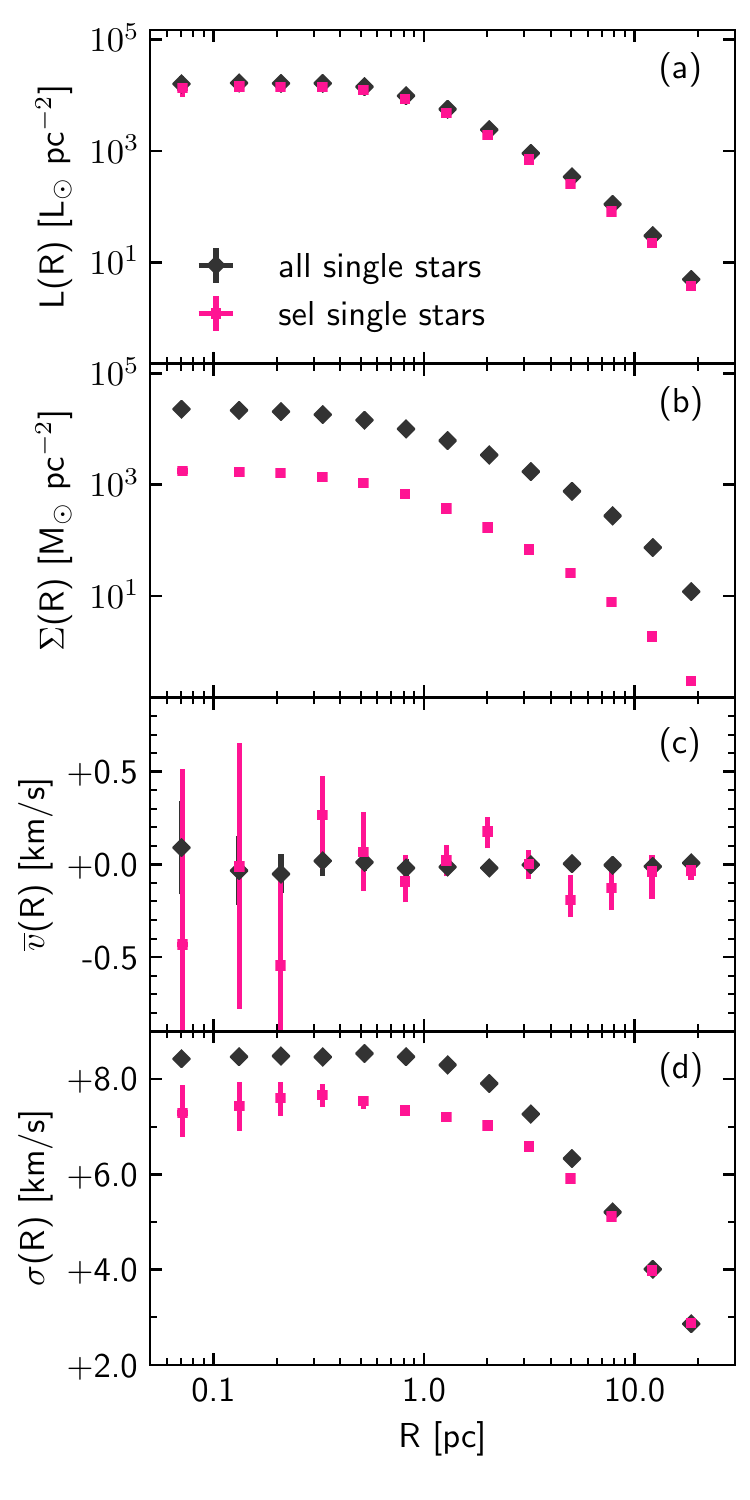}
    \caption{Radial profiles projected in the sky for the \textit{no IMBH/BHS} simulation. In panel (a), we observe no major difference on the luminosity surface density ($\text{L}(\text{R})$) between all the stars and the selected sample, this is expected as the luminosity surface density is dominated by the bright stars. This is not the case for the mass surface density ($\Sigma(\text{R})$) in panel (b) where the selection is approximately $\sim 13$ times lower than the full sample. Panels (c) and (d) shows the line-of-sight mean velocity and velocity dispersion, only in the latter we observe a $\sim10\%$ difference within $1\,R_h$ due energy equipartition effects.}
    \label{fig:PRJ_PRF}
\end{figure}
}

\item[(4)]{We fit the luminosity surface density profile given by the functional form defined in Equation \ref{eq:surf_lumn}. This allows us to cover different types of luminosity surface density profiles and  deproject them for the dynamical models. We fit the luminosity surface density with \textsc{EMCEE} \citep{foreman-mackey_2013}, a Monte Carlo Markov Chain (MCMC) sampler, which allows us to explore the multi-parameter space. From the fitting we save the best-fit parameters as input for our dynamical models. Figure \ref{fig:LUM_FIT} shows the luminosity surface brightness profiles and the fit from our MCMC approach for all the different simulations.

\begin{figure}
\begin{center}
\includegraphics[width=0.8\linewidth]{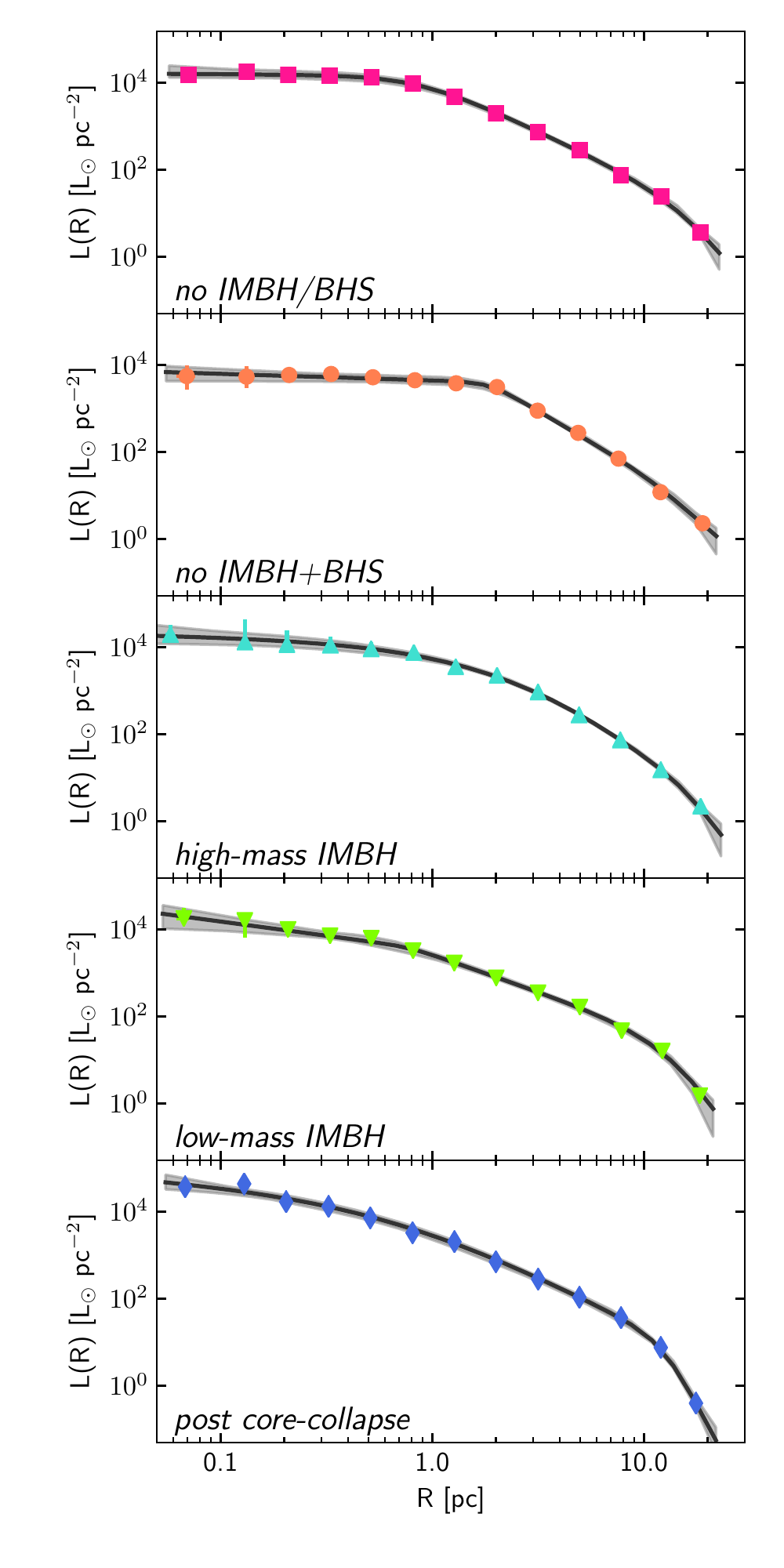}
\caption{Surface brightness profile and best fit model. For each GC we fit a functional form for the luminosity surface density as given by Equation \ref{eq:surf_lumn}. The best fit in each case (black line) will serve as the main ingredient to our dynamical models, as we assume a constant mass-to-light ratio. }
\label{fig:LUM_FIT}
\end{center}
\end{figure}
 }

\item[(5)]{We build a grid of dynamic models via the Jeans equations as described in Section \ref{sec:dyn}, based on the best-fit parameters to the surface brightness profile. Each model is defined by three parameters: the mass-to-light ratio ($\Upsilon_0$), the velocity anisotropy ($\beta$) and the mass of the central IMBH ($M_{\bullet}$).  The grid is given by the parameter space: $0.5\leq \Upsilon_0 \leq 3.5$, $-1.0 \leq \log{(M_{\bullet}/M_{\odot})} \leq 5.0$ and $-1.0 \leq \beta \leq 1.0$. For each model we calculate the Chi-square ($\chi_k^2$) as:
\begin{equation}
\chi_k^2 = \displaystyle\sum\frac{\left( \langle\text{v}_k\rangle^{1/2}_{\text{data}} -\langle\text{v}_k\rangle^{1/2}_{\text{model}} \right)^2}{\left(\delta\langle\text{v}_k\rangle^{1/2}_{\text{data}}\right)^2}\,,
\label{eq:chi2}
\end{equation}
where $k$ represent each of the observed velocities (LOS, PMR and PMT). We explore the best fit parameters first with only line-of-sight velocities, then with only proper motions and finally with all of them.
}

\end{itemize}

\subsection{Results}
\label{sec:results}
We applied the pipeline described in Section \ref{sec:pipeline} to all simulated GCs introduced in Section \ref{sec:models} and Tables \ref{tab:sim-pars-0Gyr} and \ref{tab:sim-pars-12Gyr}. Figure \ref{fig:DYN-ALL-LOS} shows our fitted dynamical models when only line-of-sight velocities (LOS) are used, while Figure \ref{fig:DYN-ALL-PMS} shows the case when radial (PMR) and tangential (PMT) proper motions are used together to constrain the best-fit parameters. Figure \ref{fig:DYN-ALL-LOS+PMS}, on the other hand, shows the results when LOS and proper motions are used together to constrain the parameters. In each figure we show the respective second velocity moment profiles ($\langle \text{v}^2 \rangle^{1/2}$) used in the $\chi^2$ minimization on the left-side panels and the parameter space on the right-side panels. We adopt three relative $\Delta\chi^2$ regions\footnote{The non-linearity and complexity of our model does not allow us to have a clear value for the degrees of freedom in our $\chi^2$ minimization. The three values adopted here represent the $1\sigma$, $2\sigma$ and $3\sigma$ for a $\chi^2$ distribution with 3 degrees of freedom. This is the case for the $\Delta\chi^2$ of a linear model with 3 free parameters.\label{foot:chi2}} given by $\Delta\chi^2 = 3.5$, $\Delta\chi^2 = 7.8$ and $\Delta\chi^2 = 11.3$ as a guide to our dynamical model and parameter distribution from the $\chi^2$ minimization. We included the best-fit parameters as an open circle on the right-side panels, while the expected values from the simulation are included as an `$\times$' (see Table \ref{tab:sim-pars-12Gyr}). For the \textit{no IMBH+BHS} simulation, we indicate with an arrow the total mass in stellar black holes within the central parsec of the cluster. Table \ref{tab:sim-fits} summarizes the best-fit parameters for all models and kinematic data, the errors in each parameter are given by the $\Delta\chi^2=7.8$ region in the figures (approximately $2\sigma$).

\begin{table}
\centering
\caption{Best-fit parameters for all the simulated GCs and velocity data used for the fits. The error bars represent the region defined by $\Delta\chi^2\leq 7.8$ (approximately $2\sigma$, see footnote \protect\ref{foot:chi2}). The first row for each GC indicates the expected values as indicated in Table \protect\ref{tab:sim-pars-12Gyr}.}
\label{tab:sim-fits}
\begin{tabular}{llccc}
\hline
Model   &  Data     &  $\Upsilon_0$ & $\log({M_{\bullet}/M_{\odot}})$ & $\beta$ \\
\hline
no IMBH/BHS&  & $1.38$ & --   & $0.03$ \\     
    & RVs & $ 1.4_{-0.55}^{+0.45}$ & $ 3.3_{-4.35}^{+0.75}$ & $-0.4_{-0.65}^{+0.55}$ \\  
    & PMs & $ 1.4_{-0.25}^{+0.25}$ & $-1.0_{-0.05}^{+4.45}$ & $-0.3_{-0.75}^{+0.35}$ \\  
    & ALL & $ 1.4_{-0.25}^{+0.25}$ & $-1.0_{-0.05}^{+4.45}$ & $-0.3_{-0.45}^{+0.35}$ \\  
\hline 
no IMBH+BHS&  & $1.39$ & --   & $0.11$ \\     
    & RVs & $ 1.5_{-0.75}^{+0.85}$ & $ 3.0_{-4.05}^{+1.25}$ & $-0.0_{-1.05}^{+0.35}$ \\  
    & PMs & $ 1.6_{-0.55}^{+0.55}$ & $ 2.7_{-3.75}^{+1.25}$ & $-0.1_{-0.95}^{+0.45}$ \\  
    & ALL & $ 1.5_{-0.35}^{+0.55}$ & $ 2.8_{-3.85}^{+1.05}$ & $-0.0_{-0.75}^{+0.25}$ \\  
\hline 
high-mass IMBH&  & $1.26$ & 4.11 & $0.10$ \\     
    & RVs & $ 1.1_{-0.65}^{+0.95}$ & $ 4.3_{-5.35}^{+0.25}$ & $-0.5_{-0.55}^{+1.25}$ \\  
    & PMs & $ 1.3_{-0.55}^{+0.65}$ & $ 4.1_{-0.25}^{+0.25}$ & $-0.0_{-1.05}^{+0.45}$ \\  
    & ALL & $ 1.2_{-0.45}^{+0.55}$ & $ 4.2_{-0.35}^{+0.15}$ & $-0.2_{-0.85}^{+0.65}$ \\  
\hline 
low-mass IMBH&  & $1.40$ & 2.72 & $0.04$ \\     
    & RVs & $ 1.4_{-0.35}^{+0.25}$ & $ 3.2_{-0.95}^{+0.25}$ & $-0.8_{-0.25}^{+0.85}$ \\  
    & PMs & $ 1.4_{-0.25}^{+0.25}$ & $ 2.5_{-3.55}^{+0.65}$ & $-0.2_{-0.85}^{+0.45}$ \\  
    & ALL & $ 1.4_{-0.25}^{+0.15}$ & $ 2.8_{-3.85}^{+0.35}$ & $-0.3_{-0.65}^{+0.45}$ \\  
\hline 
post core-collapse&  & $1.24$ & --   & $0.0$  \\     
    & RVs & $ 1.1_{-0.25}^{+0.25}$ & $ 3.0_{-4.05}^{+0.35}$ & $-1.0_{-0.05}^{+0.95}$ \\  
    & PMs & $ 1.1_{-0.15}^{+0.15}$ & $-1.0_{-0.05}^{+3.75}$ & $-0.2_{-0.55}^{+0.25}$ \\  
    & ALL & $ 1.1_{-0.15}^{+0.15}$ & $ 1.0_{-2.05}^{+1.75}$ & $-0.3_{-0.45}^{+0.25}$ \\  
\hline

\end{tabular}
\end{table}

\subsubsection{Constraints from line-of-sight velocities (LOS) only}
\label{sec:only-LOS}
\begin{figure*}
    \centering
    \includegraphics[width=0.6\linewidth]{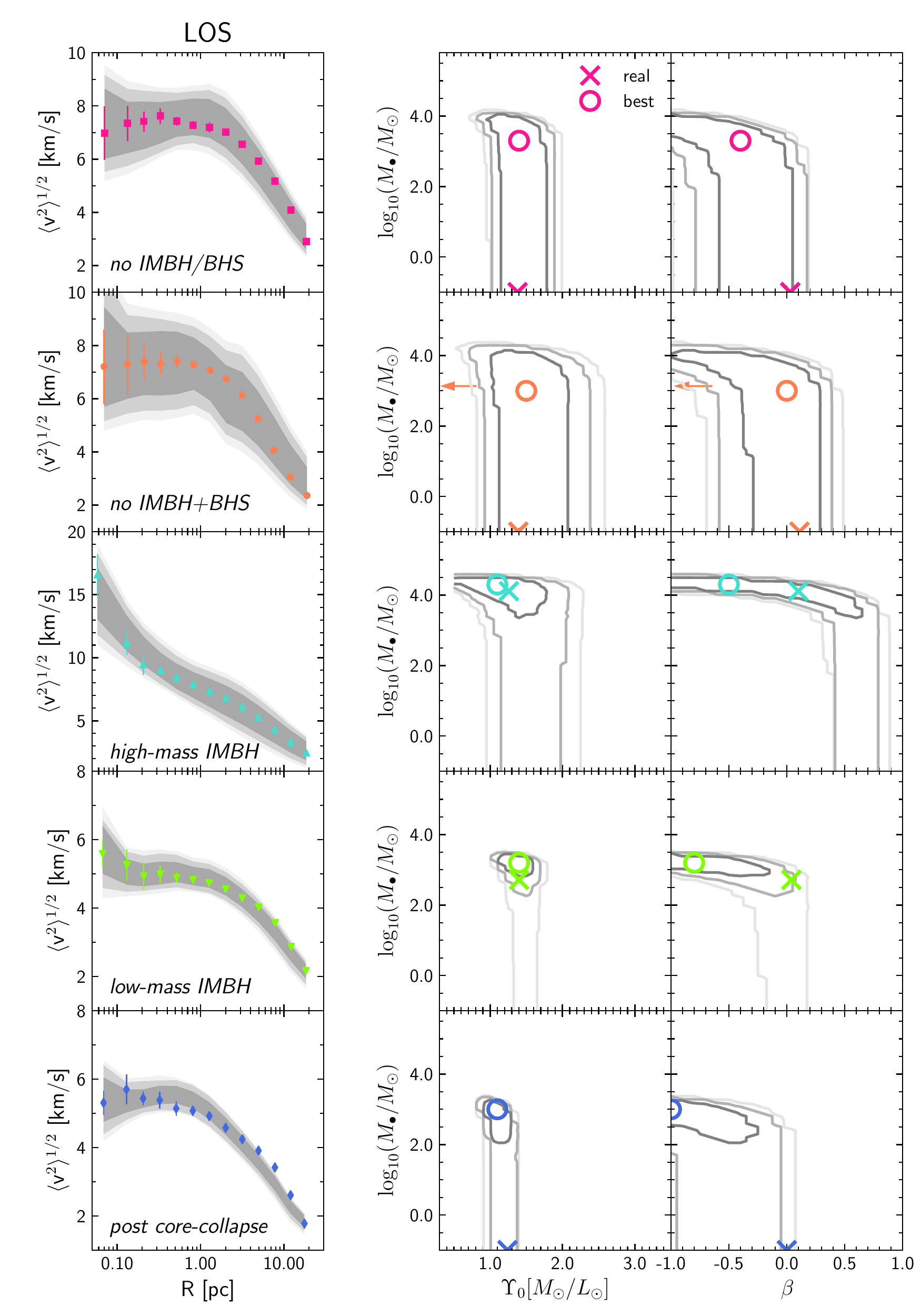}
    \caption{Fitted dynamical models and parameters space when only line-of-sight velocities (LOS) are used for the fit. The left panels show the measured second velocity moment projected in the sky (colored symbols), while the shaded are represent the $\Delta\chi^2 = 3.5$, $\Delta\chi^2 = 7.8$ and $\Delta\chi^2 = 11.3$ regions (from darker to lighter grey). The right panels shows the parameter space, whereas the circles mark the best-fit values (as in Table \ref{tab:sim-fits}) and the `$\times$' marks the expected value measured directly from the simulations (as in Table \ref{tab:sim-pars-12Gyr}), the contours represent the $\Delta\chi^2 = 3.5$, $\Delta\chi^2 = 7.8$ and $\Delta\chi^2 = 11.3$ regions. For the \textit{no IMBH+BHS} cluster, we indicate with arrows the total mass in stellar black holes (BHS) within the central $1\,\text{pc}$ of the cluster. Over all the mass-to-light ratio $\Upsilon_{0}$ is well constraint by only using LOS velocities. This is not the case for the velocity anisotropy, as the lack of constraints allows the models to have higher masses for the central IMBH at the cost of more tangential orbits. In the case of the \textit{high-mass IMBH}, the cusp in $\langle v^2 \rangle^{1/2}$ is significant enough to detect the IMBH at its centre.}
    \label{fig:DYN-ALL-LOS}
\end{figure*}

Our models can identify the presence of a central IMBH inside the two GCs which do indeed contain one (see the right side panels of Figure \ref{fig:DYN-ALL-LOS}). In the case of the \textit{high-mass IMBH} GC, our best fit value is $M_{\bullet}\sim2\pm2\times10^4\,M_{\odot}$\footnote{The quoted error bars represent the $\chi^2\leq7.8$ confidence region.}. While we obtain a detection within the $\Delta\chi^2 = 3.5$ region ($\sim 1\sigma$) which also contains the real value ($M_{\bullet}=12883.4\,M_{\odot}$), we cannot fully exclude a lower mass IMBH nor the no IMBH solution with larger confidence levels. This is likely due the lacks of constrains in the velocity anisotropy, as the parameter region with lower mass IMBHs is dominated by highly radial velocity anisotropy ($\beta\gtrsim0.5$). For the \textit{low-mass IMBH} we find a detection at $\Delta\chi^2 = 7.8$ level ($\sim 2\sigma$), where the IMBH best fit value is $M_{\bullet}\sim1.5\pm1.4\times10^3\,M_{\odot}$, around $3$ times the mass of the actual IMBH ($M_{\bullet}=519\,M_{}\odot$). This overestimation goes in hand with the high tangential anisotropy of $\beta=-0.8$, inferred from the best fit model (see discussion in Section \ref{sec:mass_constraints_ani} below).

For the \textit{no IMBH/BHS} and \textit{no IMBH+BHS} GCs we obtain upper limits of $M_{\bullet}\lesssim 11000\,M_{\odot}$ and $M_{\bullet}\lesssim 17000\,M_{\odot}$, respectively. While the whole mass range from the correct solution ($M_{\bullet} = 0\,M_{\odot}$) to the just mentioned upper limits is allowed by the model within the $\chi^2\leq7.8$ confidence region, the best fit models indicates a central IMBH of $M_{\bullet}\sim 2^{+11}_{-2}\times10^3\,M_{\odot}$ for the \textit{no IMBH/BHS} and $M_{\bullet}=1^{+17}_{-1}\times10^3\,M_{\odot}$ for the \textit{no IMBH+BHS}. Finally, although the \textit{post core-collapse} GC does not have a central IMBH, the best fit model suggests a central IMBH of $M_{\bullet} = 1_{-1}^{+1.5}\times10^3\,M_{\odot}$, which is detected within $1\sigma$. In a similar fashion than for the \textit{low-mass IMBH}, the inferred mass of the IMBH is bound to a tangential anisotropy ($\beta=-1.0$, at the edge of our parameter space).

As expected, we cannot constrain the velocity anisotropy with only LOS velocities. Figure \ref{fig:DYN-ALL-LOS} shows the existence of a correlation between the mass of the
possible IMBH and the velocity anisotropy for each of the five analyzed GCs. Dynamical models with a significant tangential anisotropy allow for a larger central IMBH mass (commonly refered to as mass-anisotropy degeneracy, see Section \ref{sec:mass_constraints_ani}). Note that for all GCs, the correlation becomes stronger for dynamical models with central IMBH masses higher than $1000\,M_{\odot}$. In all simulated GCs, we observe that our models are consistent with the observed kinematics. For the case of the \textit{no IMBH+BHS} simulation, we notice that our models overestimate the second velocity moment at  $R\gtrsim 2R_{h}$ (or $R\gtrsim 6\,\text{pc}$).

\subsubsection{Constraints from proper motions (PMs) only}
\label{sec:only-pm}
The second velocity moments for the proper motions have a different parametric dependency with the velocity anisotropy (see Equations \ref{eq:v2pmr} and \ref{eq:v2pmt}), adding an additional constraint. This improves the constraints for our models when compared with the case with only line-of-sight velocities, as the degeneracy between the velocity anisotropy and the mass of the central IMBH is reduced. Our models, however, show some limitations as when using proper motions, they become less consistent with the observed kinematics. For the \textit{no IMBH/BHS}, \textit{low-mass IMBH} and \textit{post core-collapse} GCs, the models fail to mutually fit the radial (PMR) and tangential (PMT) proper motions.

With the additional constraints provided by proper motions, we find a clear $3\sigma$ detection for the \textit{high-mass IMBH} GC and a best fit value of $M_{\bullet}\sim1.2_{-0.6}^{+1.2}\times10^4\,M_{\odot}$, which is consistent with the real mass of the central IMBH.

The best fit for the \textit{low-mass IMBH} reduces to $M_{\bullet}\sim 0.3_{-0.3}^{+1.2}\times10^3\,M_{\odot}$, which slightly underestimates the mass of the central IMBH. While we recover a best fit value which is more consistent with the real IMBH mass, we do not find a clear detection at $1\sigma$ nor $2\sigma$, the $2\sigma$ errors allow for a range of masses of $[0\,M_{\odot},1584\,M_{\odot}]$ for the central IMBH.

The constrains for the \textit{no IMBH/BHS} and \textit{no IMBH+BHS} GCs also improve. The upper limits reduces to $M_{\bullet}\lesssim 3100\,M_{\odot}$ and $M_{\bullet}\lesssim 9900\,M_{\odot}$, respectively. The best fit value for the \textit{no IMBH/BHS} is $M_{\bullet}\sim0^{+3.1}\times10^3\,M_{\odot}$, which is consistent with no central IMBH. For the \textit{no IMBH+BHS} GC simulation, the best fit is now $M_{\bullet}\sim0.5^{+9.4}_{-0.5}\times10^3\,M_{\odot}$, more consistent with the no IMBH solution. However, within $2\sigma$ it is not possible to fully rule out a higher mass IMBH.

The \textit{post core-collapse} GC also shows an improvement with a best fit IMBH mass which is consistent with zero ($M_{\bullet}\sim0.0^{+0.6}\times10^3\,M_{\odot}$). The upper limit reduces to $M_{\bullet}\lesssim 630\,M_{\odot}$ given the additional constraints on the velocity anisotropy with a recovered value of $\beta=-0.2_{-0.55}^{+0.25}$, which is closer to the actual value obtained from the simulation ($\beta_{50\%}=0.0$).

\begin{figure*}
    \centering
    \includegraphics[width=0.8\linewidth]{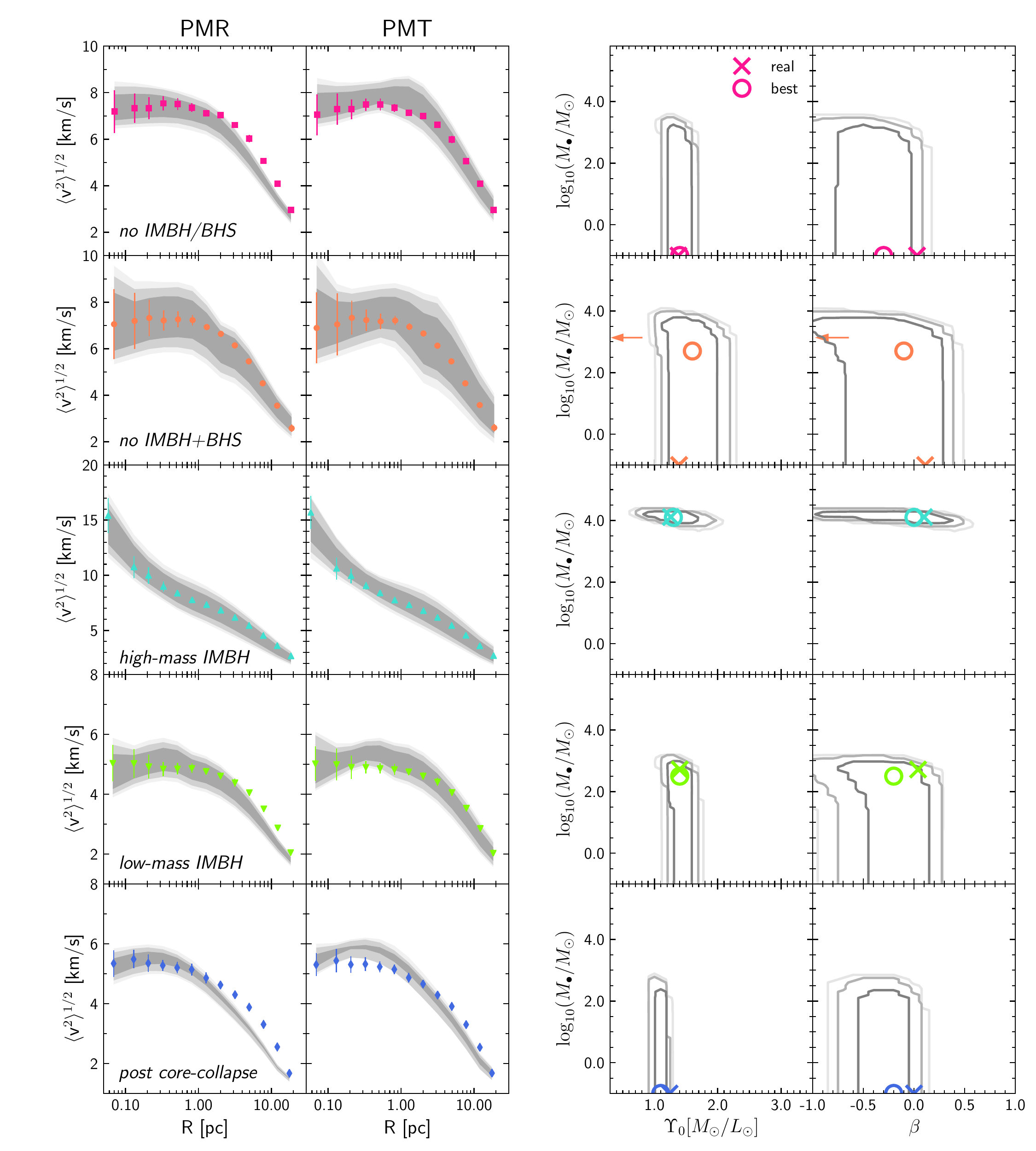}
    \caption{As in Figure \ref{fig:DYN-ALL-LOS}, but when only the proper motions (PMR and PMT) are used for the fit. The additional data allows to have a better constraint in the velocity anisotropy, excluding all the models with significant tangential orbits. Although, the constraints for the mass of the possible central IMBH are similar to when only LOS are used in the fit.}
    \label{fig:DYN-ALL-PMS}
\end{figure*}

\subsubsection{Constraints from the full kinematic sample (LOS+PMs)}
\label{sec:full-kin}
\begin{figure*}
    \centering
    \includegraphics[width=1.0\linewidth]{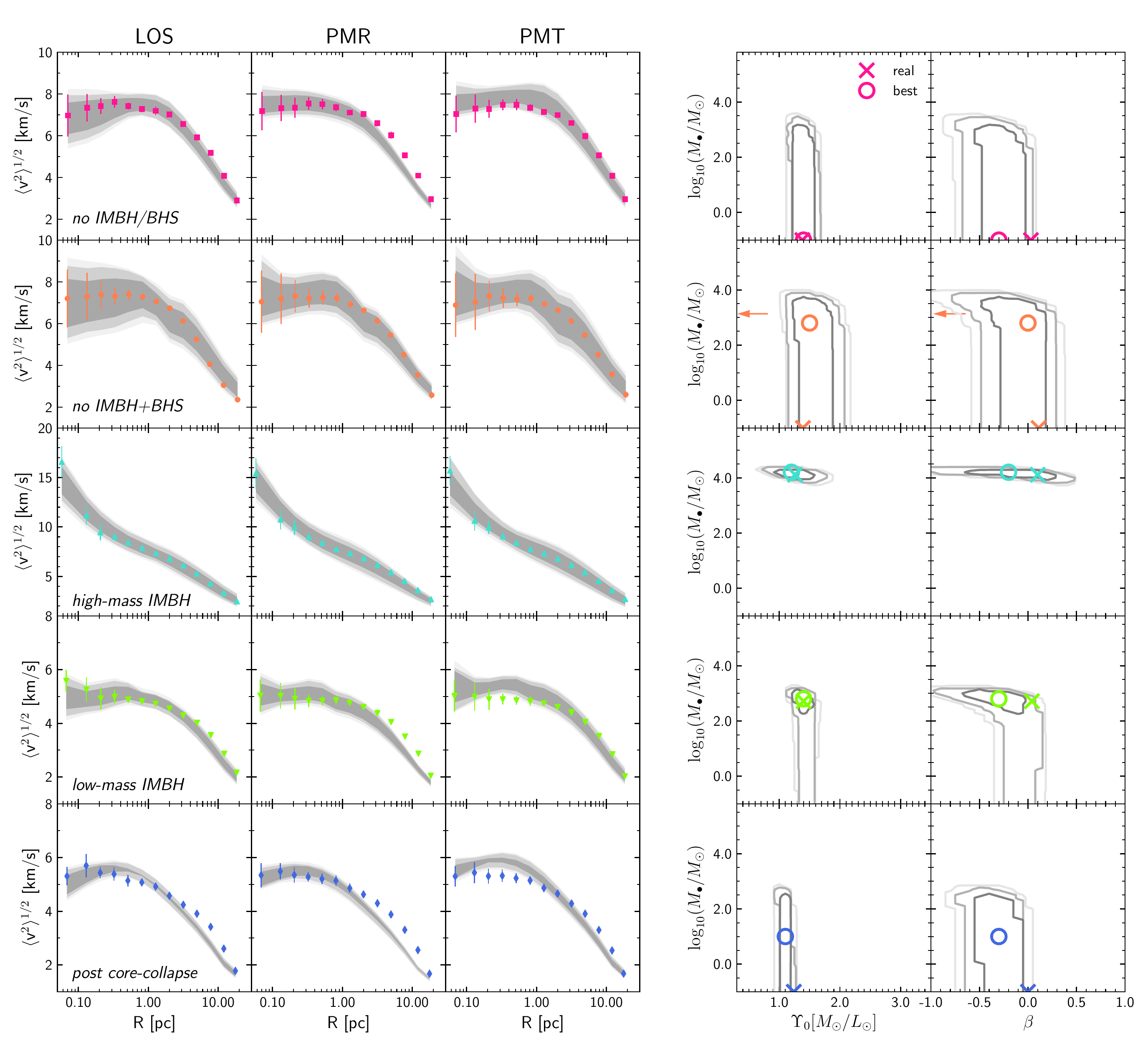}
    \caption{As in Figure \ref{fig:DYN-ALL-LOS}, but when all velocities (LOS+PMs) are used for the fit. Compared to the constraints from the only PMs case, the fits does not improve significantly when using the full 3D kinematic data. Now we have a detection for the \textit{low-mass IMBH} within the $\Delta\chi^2\leq3.5$ level ($\sim 1\sigma$). However, models without an IMBH are still allowed within the uncertainties ($\Delta\chi^2\leq7.8$ level, $\sim 2\sigma$). The upper limits on the inferred mass of the possible IMBH in the cases without one are still in the $M_{\bullet}\leq1000\,M_{\odot}$ range.}
    \label{fig:DYN-ALL-LOS+PMS}
\end{figure*}

When the full kinematic sample is used to constrain the parameter space, as shown in Figure \ref{fig:DYN-ALL-LOS+PMS}, we observe similar constraints on the different $\Delta\chi^2$ confidence regions as in the only proper motions case. The IMBH in the \textit{high-mass IMBH} GC is again clearly identified with an inferred mass of $M_{\bullet}\sim1.5\pm0.9\times10^4\,M_{\odot}$, while for the central IMBH in the \textit{low-mass IMBH} simulation we find $M_{\bullet}\sim0.6_{-0.6}^{+0.9}\times10^3\,M_{\odot}$ and its presence is recovered within $1\sigma$ level. However, for larger confidence regions, we have models that still are consistent with a lower mass or no IMBH solution.

As in the case with only proper motions, the best fit value for the \textit{no IMBH/BHS} GC is consistent with not having an IMBH ($M_{\bullet}\sim0^{+3.2}\times10^3\,M_{\odot}$), while still allowing a large upper limit ($M_{\bullet}\lesssim2800\,M_{\odot}$). Similarly, for the \textit{no IMBH+BHS} GC, we obtain an upper limit of $M_{\bullet}\lesssim 7900\,M_{\odot}$ which has improved from the only proper motion case. The best fit value is now $M_{\bullet}\sim0.6_{-0.6}^{+7.3}\times10^3\,M_{\odot}$, the range of masses covered by the $2\sigma$ level goes from $0\,M_{\odot}$ to $7900\,M_{\odot}$. Also for the \textit{post core-collapse} GC, we find a similar result as when only proper motions are used with an upper limit of $M_{\bullet}\sim630\,M_{\odot}$, while the best fit value of $M_{\bullet} = 10_{-10}^{+620}\,M_{\odot}$ is consistent with not having an IMBH.

For all clusters the global mass-to-light ratio ($\Upsilon_{0}$) is well constrained, while the velocity anisotropy ($\beta$) shows a significant improvement for all clusters with the exception of the \text{high-mass IMBH}, once the proper motions are considered (see Figure \ref{fig:ML_ANI_BEST}). In the case of the \textit{high-mass IMBH}, the velocity anisotropy does not show the same level of improvement after including the proper motions, as the Keplerian rise in velocity dispersion dominates over the velocity anisotropy in the inner kinematics. However, their inclusion allows the exclusion of highly radial anisotropic models.   

As in the case when only proper motions are considered, we notice that our models are not fully consistent with the kinematic data, this is particularly true for the \textit{post core-collapse} GC. These discrepancies are originating in the assumptions of our models and show the limitations they bring into the fitting. In the following section we discuss further how the assumptions of constant velocity anisotropy and mass-to-light ratio affect the modelling and the detection of a possible IMBH.

\subsubsection{Additional kinematic samples}

To explore the effects of our selection criteria (as described in Section \ref{sec:pipeline}) we applied the dynamical models to three additional kinematic samples. Figure \ref{fig:fits_deeper_cuts}, in the appendix, shows the constraints in the parameter space for the mass-to-light ration and mass of the possible central IMBH for two fainter magnitude cuts: $4.6\,\text{mag}$ below the main sequence turn-off, following current lower limits for precise proper motions at the cluster center \citep[][]{anderson_10,libralato_2018}, and $7.5\,\text{mag}$ below the main sequence turn-off \citep{heyl_17}, which is still only possible for proper motions outside the cluster's $R_h$, but works as an extreme hypothetical case. We do not observe any significant difference with our results for the brightest selection. We notice, though, that for the fainter magnitude cuts the best fit value for $\Upsilon_0$ increase, this is expected due to the larger fraction of low-mass stars which have a systematically larger velocity dispersion (as in Figure \ref{fig:PRJ_PRF}). The third case we explored includes long period binaries ($P>1\,\text{yr}$) as in panel (b) of Figure \ref{fig:BIN_CENTER}. The comparison with our main results is illustrated in Figure \ref{fig:fits_with_binaries} and we, once again, do not observe any significant difference between our main results and the sample including long period binaries, which is also expected as both kinematic samples are similar (see Figure \ref{fig:delta_vdisp}).

\begin{figure}
    \centering
    \includegraphics[width=1.0\linewidth]{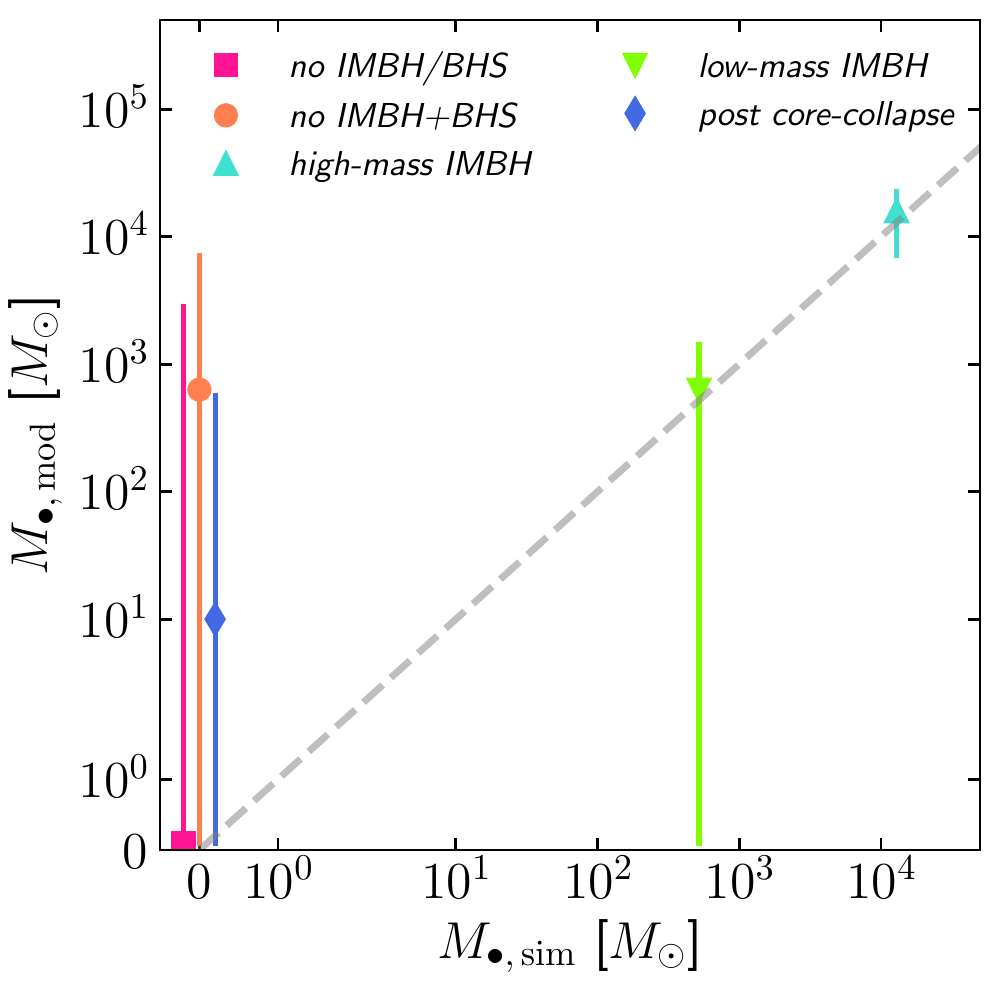}
    \caption{Recovered IMBHs masses from the full 3D kinematic sample (LOS+PMs). Our dynamical models robustly identify the IMBH in the \textit{high-mass IMBH} simulation ($M_{\bullet}/M_{GC}=4.1\%$). However,  lower masses or the absence of the central IMBH cannot be excluded for the \textit{low-mass IMBH} case ($M_{\bullet}/M_{GC}=0.3\%$). The three simulations without a central IMBH show large upper limits (with an offset from $M_{\bullet,\,sim}=0.0\,M_{\odot}$ for visibility).}
    \label{fig:imbh_masses}
\end{figure}

\section{Mass constraints from the Jeans models}
\label{sec:mass_constraints}
\begin{figure}
    \centering
    \includegraphics[width=1.0\linewidth]{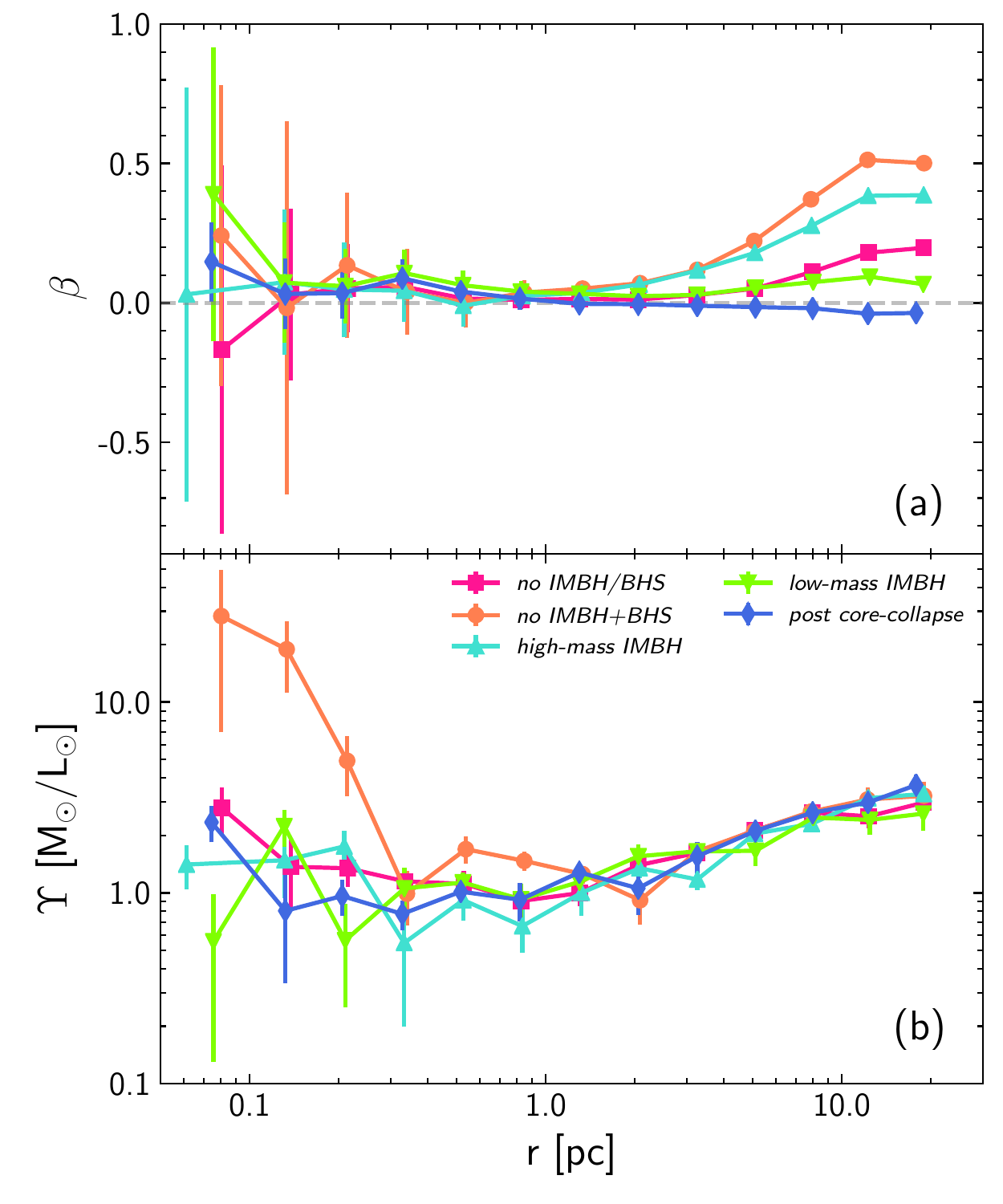}
    \caption{Velocity anisotropy (a) and mass-to-light ratio (b) profiles for each simulation. All simulated clusters, with the exception of the \textit{post core-collapse}, have central velocity anisotropies consistent with being isotropic ($\beta=0$) and become more radially anisotropic at large radii. The \textit{post core-collapse} cluster is fairly isotropic at all radii. The stellar mass-to-light ratio ($\Upsilon$) in the simulations varies with radius, increasing towards the centre and the outer regions of the cluster. The central slope of $\Upsilon$ varies with each cluster, where the \textit{no IMBH+BHS} shows the most significant increase due the stellar black holes subsystem at is centre. On the other hand all simulated GCs shows the same behaviour at large radii.}
    \label{fig:internal_ml_ani}
\end{figure}

The two main assumptions in our dynamical models, which could impact in the determination of the presence of an IMBH and its mass, are firstly the constant mass-to-light ratio and secondly the constant velocity anisotropy (see Section \ref{sec:dyn}). As shown in Figure \ref{fig:internal_ml_ani}, the internal velocity anisotropy and mass-to-light ratio vary for all five GC simulations. The velocity anisotropy increases at large radii for all GCs, other than the \textit{post core-collapse}. The mass-to-light ratio increases towards the centre and at large radii. While the central mass-to-light ratio depends on the type of central object in the cluster, the rise at large radii is similar for all simulations. In this section we explore in detail the effects of these factors on our dynamical models.

\subsection{Velocity anisotropy}
\label{sec:mass_constraints_ani}
The amount of velocity anisotropy in the central region of the GC can affect the measured mass of the possible central IMBH. A radial velocity anisotropy ($\beta>0$) at the centre can reproduce an increase of the velocity dispersion without requiring additional mass (i.e. an IMBH). On the other hand if the central anisotropy becomes more tangential ($\beta<0$) the model will require an additional mass in the centre of the GCs. This mass-anisotropy degeneracy is well known in dynamical models based on Jeans equations \citep[see][for example]{binney_1982}.

The velocity anisotropy can be constrained by including 3D kinematic data namely proper motions, as discussed in Section \ref{sec:results}. However, how strongly the anisotropy can be constrained will depend on the quality of the available proper motions. In the case of NGC 5139, \cite{van_der_marel_2010} show that anisotropic models are necessary to describe its observed kinematics and provide good fits to the observed proper motions without the need for a central IMBH, when using models based on Jeans equations. More recently, \cite{zocchi_2017} also show that models based on anisotropic distribution functions are consistent with the available kinematics of NGC 5139, while their models do not rule out a central IMBH, they put a cautionary note on the estimated mass of the central IMBH. Both works find a velocity anisotropy profile which is (or close-to) isotropic in the centre. However, while \citet{van_der_marel_2010} find a tangential anisotropy at large radii, \citet{zocchi_2017} find a radially biased anisotropy profile at large radii (before becoming once again isotropic at the tidal radius). The latter is consistente with \citet{watkins_15}, who show that most galactic globular clusters in the HSTPROMO sample are isotropic towards the centre and become radially anisotropic at large radii. The upper limit on the possible IMBH mass in NGC 5139 suggests a mass-fraction of $M_{\bullet}/M_{GC} < 0.43\%$ \citep{van_der_marel_2010} similar to our \textit{low-mass IMBH} case ($M_{\bullet}/M_{GC} = 0.30\%$). In this regime, the kinematic signature of the IMBH on the observed velocity dispersion profile is not strong enough for a clear detection and it can be reproduced as well by mildly radial anisotropic models ($\beta\sim0.1$). 

Panel (a) of Figure \ref{fig:internal_ml_ani} shows the velocity anisotropy for all five GCs measured directly from the simulations. The low number of stars in the central bins is accounted for with the error bars (through bootstrapping in each bin). All GCs except for the \textit{post core-collapse} are consistent with being isotropic at their centre and become more radially anisotropic at larger radii, while the \textit{post core-collapse} is consistent with being isotropic at almost all radii. Once we include the proper motions in our dynamical models, the fits become consistent with an isotropic velocity anisotropy ($\beta=0$, see Figures \ref{fig:DYN-ALL-PMS} and \ref{fig:DYN-ALL-LOS+PMS}), while still allowing for models with a more tangential anisotropy (within our error bars). The bias toward tangential anisotropy seems to be a common limitation of standard Jeans modelling approaches \cite[e.g. see][]{read_2017}.

\begin{figure}
    \centering
    \includegraphics[width=1.0\linewidth]{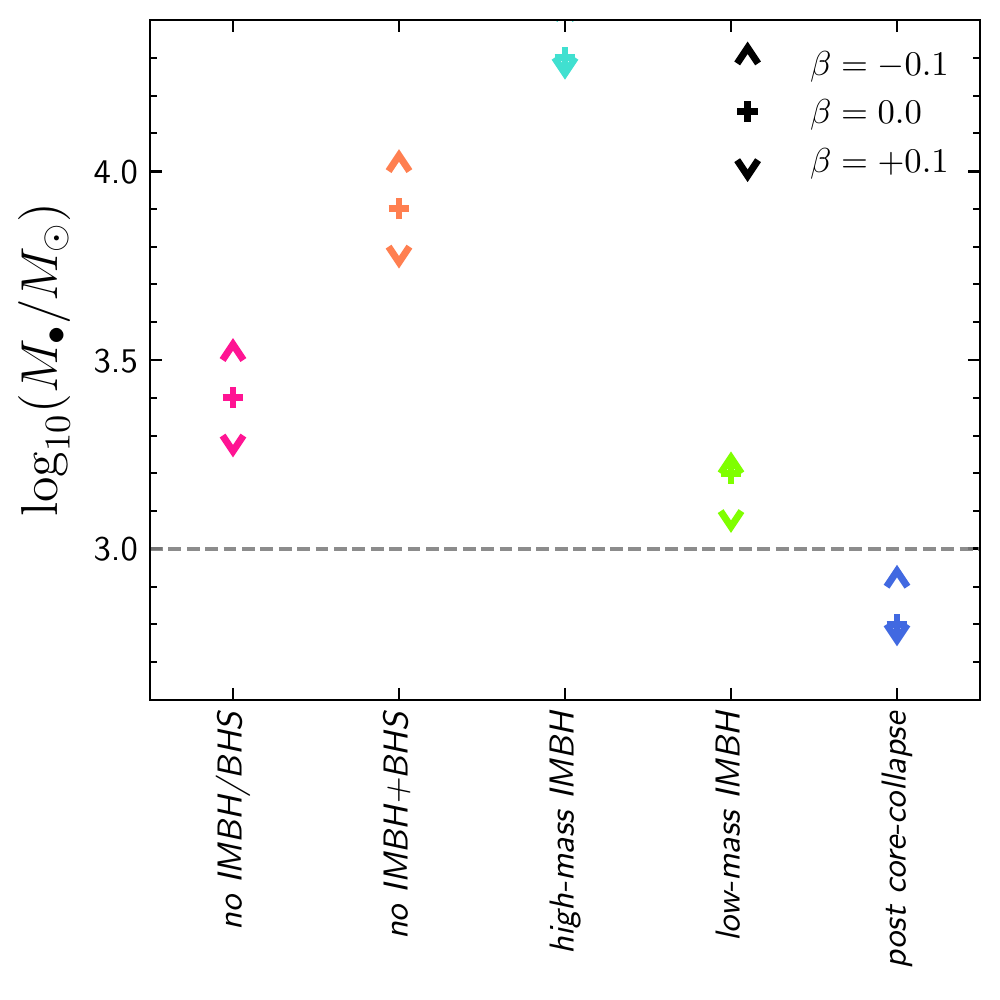}
    \caption{Upper limits of the $\chi^2\leq7.8$ region for the central IMBH mass given different velocity anisotropies for the full kinematic data case (LOS+PMs). The tangentially anisotropic case ($\beta=-0.1$, up-red arrow) gives systematically higher upper limits than the isotropic case ($\beta=0.0$, black crosses) for the inferred mass of the IMBH. On the other hand the radial anisotropic case ($\beta=+0.1$, down-blue arrow) has systematically lower upper limits, as radial anisotropy can mimic an increase of velocity dispersion in the centre (mass-anisotropy degeneracy).} 
    \label{fig:upper_mbh_ani}
\end{figure}

Figure \ref{fig:upper_mbh_ani} shows the effects of anisotropy in the upper limits of the inferred mass of the central IMBH. Models with a fixed tangential anisotropy ($\beta=-0.1$) increase the inferred IMBH mass, while models with radial anisotropy ($\beta=0.1$) reduce the upper limit. However, given the constraints from the proper motions, the variation on the upper limit of the inferred IMBH mass due anisotropy is not able to exclude the IMBH solution for the cases without one. The upper limits are still above $M_{\bullet}\sim 1000\,M_{\odot}$ ($M_{\bullet}\leq630\,M_{\odot}$ for the \textit{post core-collapse} GC).

\subsection{Mass-to-light ratio}
\label{sec:mass_constraints_ml}
As shown in panel (b) of Figure \ref{fig:internal_ml_ani}, the mass-to-light ratio of all simulations is generally not constant. The variation with radius is a direct consequence of the two-body relaxation process of collisional systems such as GCs and it has been systematically observed in simulations \citep{bianchini_2017,baumgardt_2017}, which in turn has an impact on the mass profiles of our simulated clusters and the constrains from our models.

\begin{figure}
    \centering
    \includegraphics[width=1.0\linewidth]{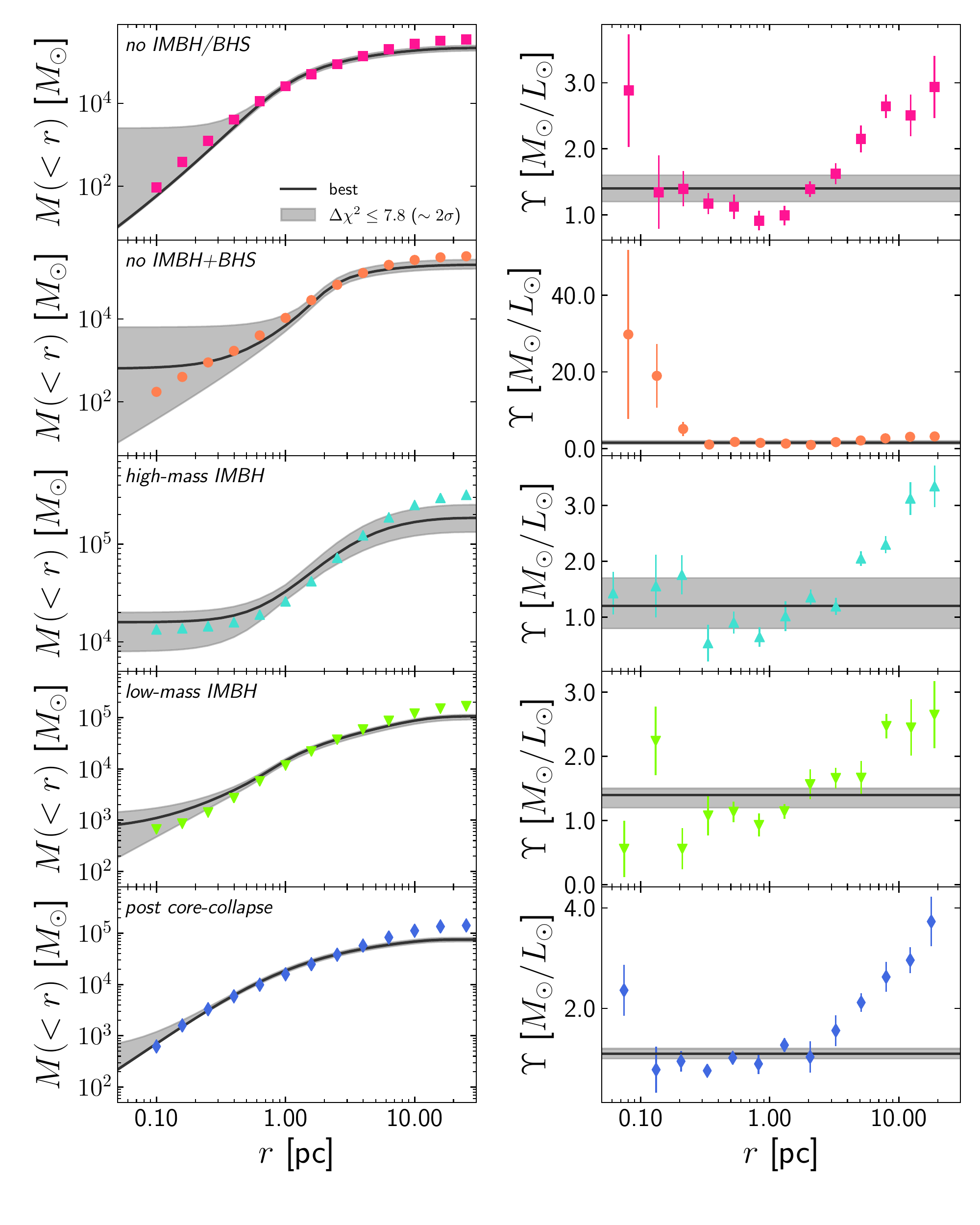}
    \caption{Mass profiles for all the simulated GCs. In the right column we include the cumulative mass profiles for each simulated GCs as coloured symbols. The black line represents the best-fit model, when all the velocity data are included in the fit. While the grey shaded area represents the $\Delta\chi^2\leq7.8$ region. The models tend to be less constrained towards the centre, in particular for the \textit{no IMBH/BHS}, \textit{no IMBH+BHS} and \textit{low-mass IMBH cases}. The right panels show the mass-to-light ratio for each simulation. These profiles differ significantly from the assumption of a constant mass-to-light ratio. The case of the \textit{no IMBH+BHS} simulation is quite extreme as the cluster of stellar-mass IMBH significantly increases the central values of the mass-to-light ratio profile. This is also shown in the cumulative mass profile, where it rises towards the centre instead of declining as in the \textit{no IMBH/BHS} or \textit{post core-collapse} simulations.}
    \label{fig:mass_profile}
\end{figure}

Figure \ref{fig:mass_profile} shows the cumulative mass profiles ($M(<r)$, left side panels) and mass-to-light ratios ($\Upsilon$, right panels) for all five simulated GCs. The shaded area represents the models with $\Delta\chi^2\leq 7.8$, while the black line represent the best fit model (for the full kinematic sample, i.e. LOS+PMs as in Section \ref{sec:full-kin}); the symbols correspond to the measured values from each simulation. For the \textit{no IMBH/BHS} and \textit{no IMBH+BHS} simulations, the central mass of the GC is poorly constrained. The value of $\Upsilon_0$ underestimates the central mass-to-light ratio of the cluster as shown in the right side panel of Figure \ref{fig:mass_profile}. The dynamical model then requires additional mass to generate the observed velocity dispersion towards the centre, allowing for the presence of an IMBH. This effect is evident in the \textit{no IMBH+BHS} case, as the cluster of stellar mass black holes increases drastically the mass-to-light ratio toward its centre. For this case the inferred mass of the central IMBH is $M_{\bullet}=631_{-631}^{+7312}\,M_{\odot}$ when using the full kinematic sample. While no false central IMBH is detected, we cannot exclude it either, as the upper limit for such an inferred central IMBH is $M_{\bullet}<7943\,M_{\odot}$. On the other hand, the presence of a central IMBH will quench mass segregation \citep[see][]{gill_2008} and in turn change the shape of the mass-to-light ratio profile. This is the case of the \textit{high-mass IMBH} simulation, where the central mass-to-light ratio is well represented by the assumption of a constant mass-to-light ratio (see Figure \ref{fig:mass_profile}).    

\begin{figure}
    \centering
    \includegraphics[width=0.8\linewidth]{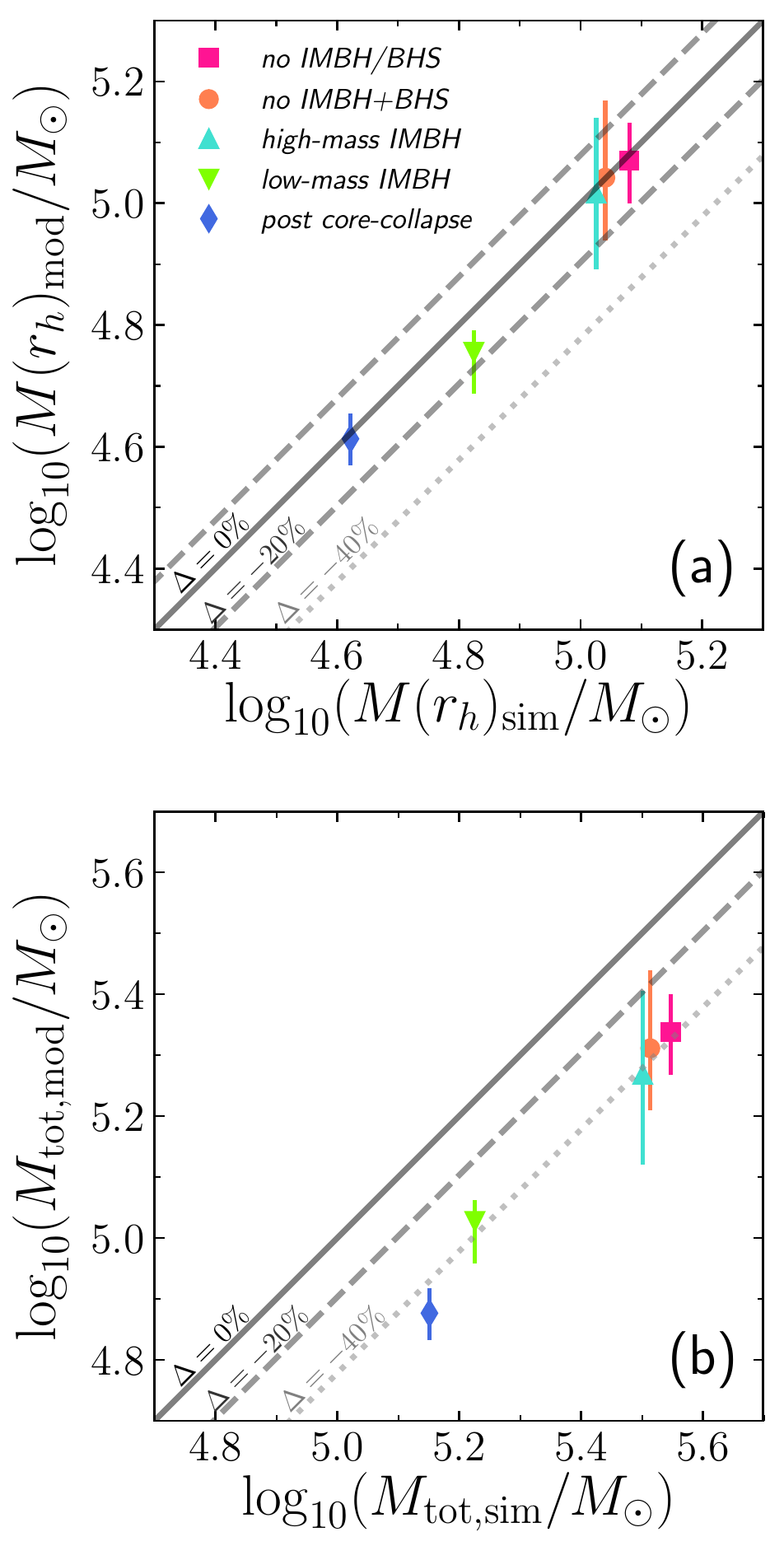}
    \caption{Recovered enclosed mass for the five simulated GCs. Panel (a): the mass within deprojected half-light radius $r_h$ is recovered for all GCs (less than $20\%$ error). On the other hand, in panel (b), the total mass of the simulated GCs is systematically underestimated.}
    \label{fig:Tot_rh_mass}
\end{figure}

The assumption of constant mass-to-light ratio is not only relevant for the central region of the simulated GCs. As massive particles sink towards the centre, the lighter ones populate the outer regions of the GC. This process also increases the mass-to-light ratio at larger radii, as faint low-mass stars dominate the exterior regions of the cluster. In panel (b) of Figure \ref{fig:internal_ml_ani} we can see that all five simulated GCs have a similar increase in their deprojected mass-to-light ratio profiles at larger radii. In the same way as for the centre of the cluster, our models underestimate the mass-to-light ratios and therefore the mass profiles (see Figure \ref{fig:mass_profile}), which in turn could bias the estimates on the cluster mass. Panel (a) of Figure \ref{fig:Tot_rh_mass} shows the recovered enclosed mass within the deprojected half-light radius $r_h$ from our dynamical models. For all five simulations our estimated mass within $r_h$ is consistent with the mass measured directly from the simulation, our fitted values for $\Upsilon_0$ are in agreement with the expected mass-to-light ratio within $r_{50\%}$ ($\Upsilon_{50\%}$, see Tables \ref{tab:sim-fits} and \ref{tab:sim-pars-12Gyr} respectively). However, this is not the case at larger radii; panel (b) in Figure \ref{fig:Tot_rh_mass} shows that for all simulated GCs their total masses are within $20\%$ and $40\%$ lower than the expected one. This is in agreement with other works: the effect of mass segregation on the recovering of global properties of GCs was discussed previously by \citet{sollima_2015}, where they applied different modelling techniques from multi-mass distribution functions to \textit{N}-body simulations of GCs. They show that single mass models systematically underestimate the total mass of the cluster, and found that the global parameters are well constrained within the radial range $r_h/2 < r < r_h$. In agreement with this, our models have a lower discrepancy on the recovered mass for radii close to $r_h$ (see Figure \ref{fig:mass_ml_error}).          

From the discussion above, one can infer that the assumption of a constant mass-to-light ratio has a larger impact on the constrains for the mass profiles, and in turn on the IMBH masses, than the assumption of constant velocity anisotropy. To characterize the real effect of these assumptions it is necessary to design a model which includes the variations on the mass-to-light ratio and velocity anisotropy profiles, which is beyond the scope of this paper.

\section{Summary}
\label{sec:summary}

The presence of IMBHs at the centre of galactic GCs is still an ongoing debate. Even with the diverse literature available on the topic \citep[][to name a few]{noyola_2008,van_der_marel_2010,lutzgendorf_2011,kamann_2014,kamann_2016,kiziltan_2017}, a robust evidence is still missing. Limitations on the observations \citep[such as kinematic centre and crowding, see][]{noyola_2010,lanzoni_2013,de_vita_2017} or in the modelling \citep[due to anisotropy or a dark component, see][]{van_der_marel_2010,zocchi_2017,zocchi_2019,mann_2019,baumgardt_2019} make the detection of IMBHs challenging. Here we explored the limitations of the dynamical model commonly used, namely models based on the Jeans equations. Using five Monte Carlo simulations of GCs with and without central IMBH from the MOCCA-survey (see Section \ref{sec:models}), we have analyzed the reliability and limitations of spherically symmetric Jeans models (see Section \ref{sec:dyn}) under the assumption of constant mass-to-light ratio and velocity anisotropy. We extracted a kinematic sample from the simulated GCs, excluding all binary systems and selecting stars brighter than $1$ magnitude below the main sequence turn-off (see Section \ref{sec:pipeline}). We fit the Jeans models to the second velocity moment profiles, varying the mass-to-light ratio ($\Upsilon_0$), the mass of the central IMBH ($M_{\bullet}$) and velocity anisotropy ($\beta$); we do so for only line-of-sight velocities (LOS, Section \ref{sec:only-LOS}), only proper motion velocities (PMs, radial and tangential on the sky, see Section \ref{sec:only-pm}) and the full kinematic sample (i.e. LOS+PMs, in Section \ref{sec:full-kin}). 

Our dynamical models can recover the mass of the high-mass IMBH ($M_{\bullet}/M_{GC} = 4.1\%$) quite well (see Section \ref{sec:results}). The kinematic signature of such an IMBH is strong and the rise in velocity dispersion cannot be explained otherwise. On the other hand for the low-mass IMBH ($M_{\bullet}/M_{GC} = 0.3\%$) we can identify the central IMBH only within $1\sigma$ (i.e. $\Delta\chi^2\leq3.5$) level, and while the best fit model is consistent with the actual mass of the central IMBH ($M_{\bullet}=519.3\,M_{\odot}$), models with no IMBH are possible within the errors (note that we only consider kinematic errors due to stochasticity of low numbers of stars per bin, observational errors could increase the uncertainty of the central IMBH mass). For all three simulations without a central IMBH we only get upper limits and while the no IMBH solution is within the range of masses, such upper limits allow for a possible IMBH in their centres.

The dynamical models are limited by two main assumptions: constant velocity anisotropy and constant mass-to-light ratio. Both have different consequences on the upper limits and detection of the central IMBH (see Section \ref{sec:mass_constraints}). Depending on the inferred amount of velocity anisotropy at the centre of the cluster, the dynamical model can slightly change the required IMBH mass to match the observed kinematics. This is relevant for identifying low-mass IMBHs. The upper limits for the inferred mass of the possible IMBH in NGC 5139 \citep{van_der_marel_2010} suggest a mass fraction of $M_{\bullet}/M_{GC} \leq 0.43\%$, which is close to our low-mass case ($M_{\bullet}/M_{GC}= 0.3\%$). While both, \citet{van_der_marel_2010} and \citet{zocchi_2017} find that anisotropic models are better when compared to the observed velocity dispersion of NGC 5139, the models by \citet{van_der_marel_2010} do not require a central IMBH to explain its observed kinematics. On the other hand, \citet{zocchi_2017} suggest strict upper limits, but do not rule out a central IMBH. Better understanding of the velocity anisotropy profiles and the effects of velocity errors on the analysis are necessary to fully disentangle the effects of anisotropy on the inferred mass of low-mass IMBHs. For the cases without an IMBH, we observe that anisotropy alone cannot reduce the upper limits as including the full kinematics sample (LOS+PMs) limits the range of anisotropy that the data allows (see Figure \ref{fig:upper_mbh_ani} and Section \ref{sec:mass_constraints_ani}).

The assumption of constant mass-to-light ratio has a more significant impact on our analysis, as the mass-to-light ratio increases towards the centre and at larger radii (see panel (b) of Figure \ref{fig:internal_ml_ani}). For the cases without IMBH we underestimate the central mass due to mass segregation effects (i.e. rise in mass-to-light ratio), which allows the dynamical model to include a central IMBH to recover the observed velocity dispersion. This is even more relevant when the stellar black hole retention is higher, such as the case of the model with a stellar black hole subsystem (\textit{no IMBH+BHS}). By applying a multi-mass model which allows for a population of stellar mass black holes at the centre of NGC 5139, \citet{zocchi_2019} show that the population of black holes can reproduce the observed kinematic data, although it cannot discard completely a less massive IMBH. Using a different approach, \citet{baumgardt_2019} also show that the presence of a cluster of stellar mass black holes can explain the observed kinematics of NGC 5139. In their case, they compare the observed kinematics to a library of \textit{N}-body simulations, which intrinsically include a variable mass-to-light ratio.

The assumption of constant mass-to-light ratio not only limits our knowledge of the central mass of the GCs, but also its total mass. As two-body relaxation pushes outwards the faint low-mass stars, the mass-to-ligth ratio increases at large radii. We systematically underestimate the mass-to-light ratio in the cluster outskirts and therefore its total mass, as shown in Figure \ref{fig:Tot_rh_mass}, is systematically underestimated with a difference of $\sim 40\%$ with respect to the expected mass for all simulated clusters. We are able to recover the mass enclosed within the half-light radius, which is consistent with the radial range proposed by \citet{sollima_2015} for estimating global properties of GCs with multi-mass distribution functions. Further improvements to our Jeans code are necessary to investigate if we can solve these issues by relaxing the constant mass-to-light ratio assumption.

GCs are collisional systems and their dynamical evolution is tied to the two-body relaxation process. Therefore, it is necessary to include the effects of collisionality in the dynamical models to be able to explain the observed kinematics, even more to robustly identify IMBHs at the centre of GCs. The results of applying our models to the \textit{high-mass IMBH} ($M_{\bullet}/M_{GC}=4.1\%$) suggest that there is a mass-fraction limit where the effects of collisionality can be excluded from the analysis, finding this limit requires further investigation beyond the scope of this paper. Ultimately, this will help to understand where we must improve the dynamical models. Most GC candidates for having an IMBH are in the low-mass range with  $M_{\bullet}/M_{GC} \lesssim 1.0\%$ \citep{van_der_marel_2010}, where the kinematic signature can also be explained by the effects of collisionality such as mass segregation, energy equipartition and a variable mass-to-light ratio. To be able to disentangle the different sources of a velocity dispersion rise in the centre of GC, models that can describe properly the mass profile of a GCs are a must. Recently, \citet{henault-brunet_19} provide a compilation of different dynamical methods and their reliability for recovering GC properties. Methods with multiple mass populations and variable mass-to-light ratio significantly improve the recovery of the mass profiles of GCs, although are still limited by observational constraints and large error bars.

While observational limitations will further complicate the detection of IMBHs in GCs, we have taken the first step in better understanding the ability to recover an IMBH from data with models based on the Jeans equation. The limitations presented here are identical for any such model under the same assumptions, not just ours. While the dynamical models studied here do not lead towards a biased solution, they lack the sensitivity to robustly infer the presence or absence of a low-mass IMBH. Improving a model’s ability to recover the mass profiles of GCs, and further understanding how the constant mass-to-light and velocity anisotropy assumptions along with the observed kinematics influence a model is crucial towards robustly identifying or rule out the presence of IMBHs in galactic GCs. We will further address observational challenges such as binaries in a subsequent paper.


\section*{Acknowledgements}
We thank the anonymous referee for their constructive comments which greatly improved this manuscript. We thank the \textsc{MOCCA}-Survey collaboration for making their data available to us and answering all our questions. We thank Nadine Neumayer, Laura Watkins and Manuel Arca Sedda for useful discussions. FIA and GvdV acknowledges funding from the European Research Council (ERC) under the European Union's Horizon 2020 research and innovation programme under grant agreement No 724857 (Consolidator Grant ArcheoDyn). FIA acknowledges funding from DAAD PPP project number 57316058 "Finding and exploiting accreted star clusters in the Milky Way" for a collaboration visit. ACS is supported by the Deutsche Forchungsgemeinschaft (DFG, German Research Fundation) -- Project-ID 138713538 -- SFB 881 ("The Milky Way System", subproject A08), which also provided PB and AA with funding for a collaboration visit. AMB acknowledges support by the same SFB 881 grant. AA acknowledges support from the Carl Tryggers Foundation for Scientific Research through the grant CTS 17:113 and from the Swedish Research Council through the grant 2017-04217.

This research made use of the \textsc{NUMPY} package \citep{van_der_walt_2011}, while all figures were made using \textsc{MATPLOTLIB} \citep{hunter_2007}.  

\section{Data availability}
The simulated GCs data underlying this article were provided by the \textsc{MOCCA} group\footnote{\url{https://moccacode.net/}} by permission. The data will be shared on request to the corresponding author with permission of the \textsc{MOCCA} group. The code to solve the Jeans equations and generate the dynamical models presented in this article will be shared on request to the corresponding author.




\bibliographystyle{mnras}
\bibliography{paper_01} 




\appendix

\section{Additional figures}

\begin{figure*}
    \centering
    \includegraphics[width=0.7\linewidth]{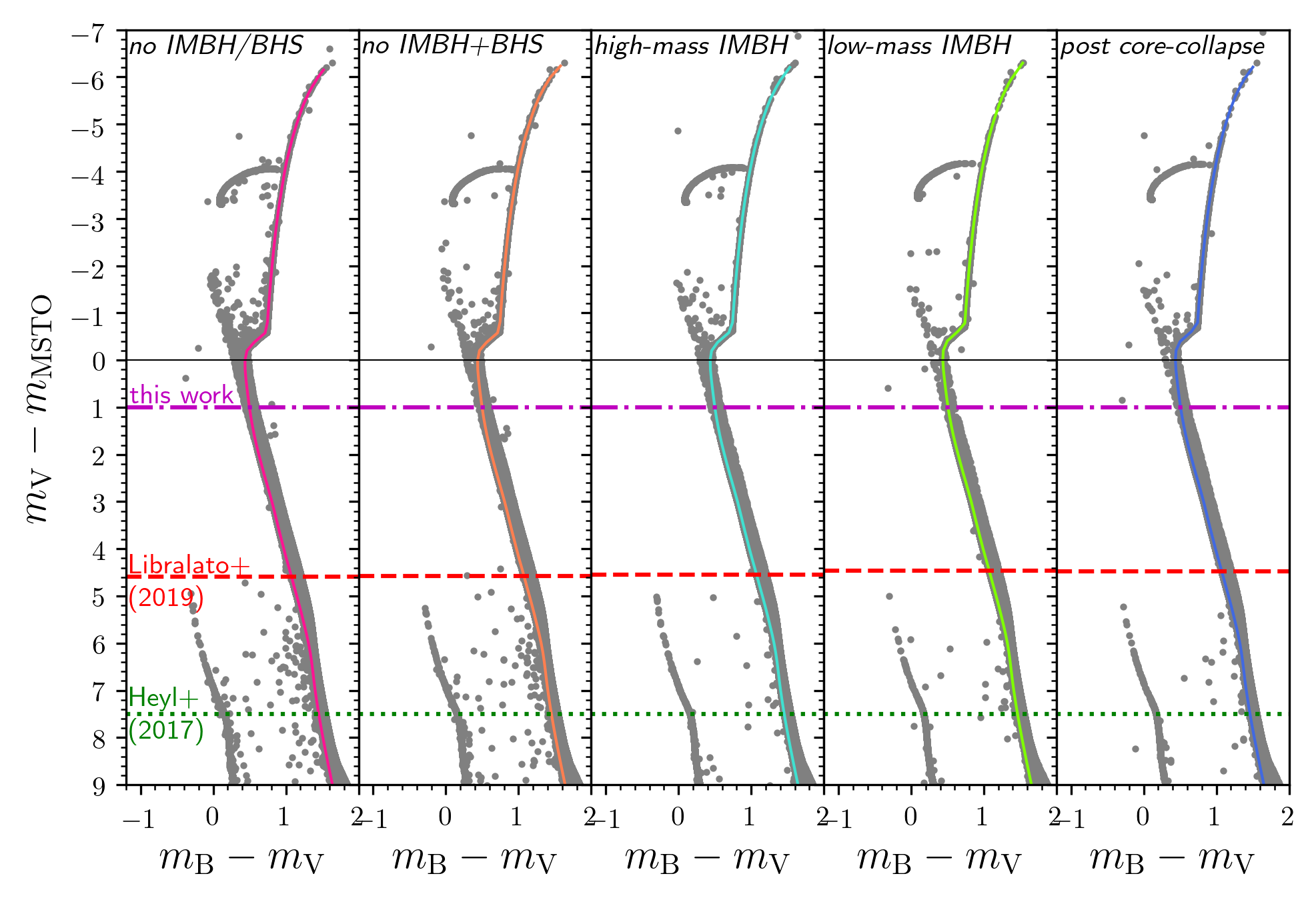}
    \caption{CMD for all five GC simulations, each of them centred at their respective MSTO magnitude. Our selection on magnitude is represented by the dot-dashed line and it is equivalent to select all stars brighter than $m_V \sim 18.5$ at a distance of $5\,\text{kpc}$ (as described in Section \ref{sec:pipeline}) and follows the magnitude limit in \protect\cite{watkins_15} for HST proper motions. For comparison we include limits from HST data for the central \protect\citep[][for NGC 362]{libralato_2018} and outer \protect\citep[][for NGC 104]{heyl_17} regions of a GC.}
    \label{fig:msto_lum_cut}
\end{figure*}

\begin{figure*}
    \centering
    \includegraphics[width=0.5\linewidth]{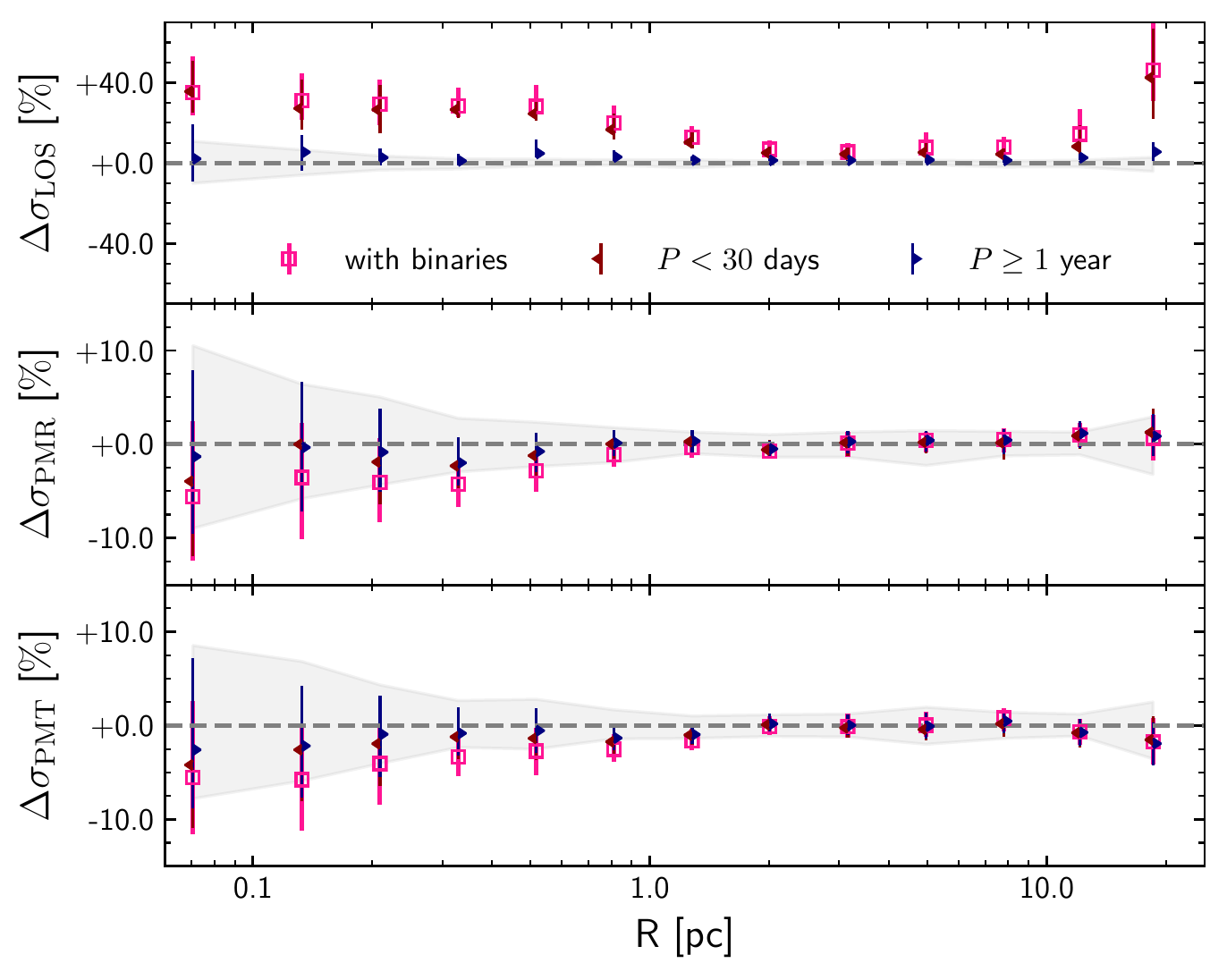}
    \caption{Difference in velocity dispersion for different binary populations relative to the sample without binaries, for the \textit{no IMBH/BHS} simulation (as in Figure \ref{fig:BIN_CENTER}). Binary systems have different effects in the velocity dispersion for each type of kinematic data. The observed line-of-sight (LOS) velocity of binary systems is mostly dominated by their internal orbital velocity, which translate in a increase in the measured velocity dispersion and it is mostly dominated by short period binaries ($P<30\,\text{days}$). On the other hand, proper motions (radial (PMR) and tangential (PMT) components) are not affected by the internal orbital motion of each component, rather the measured velocity dispersion will be affected by the level of energy equipartition of the binary systems.}
    \label{fig:delta_vdisp}
\end{figure*}

\begin{figure*}
    \centering
    \includegraphics[width=0.8\linewidth]{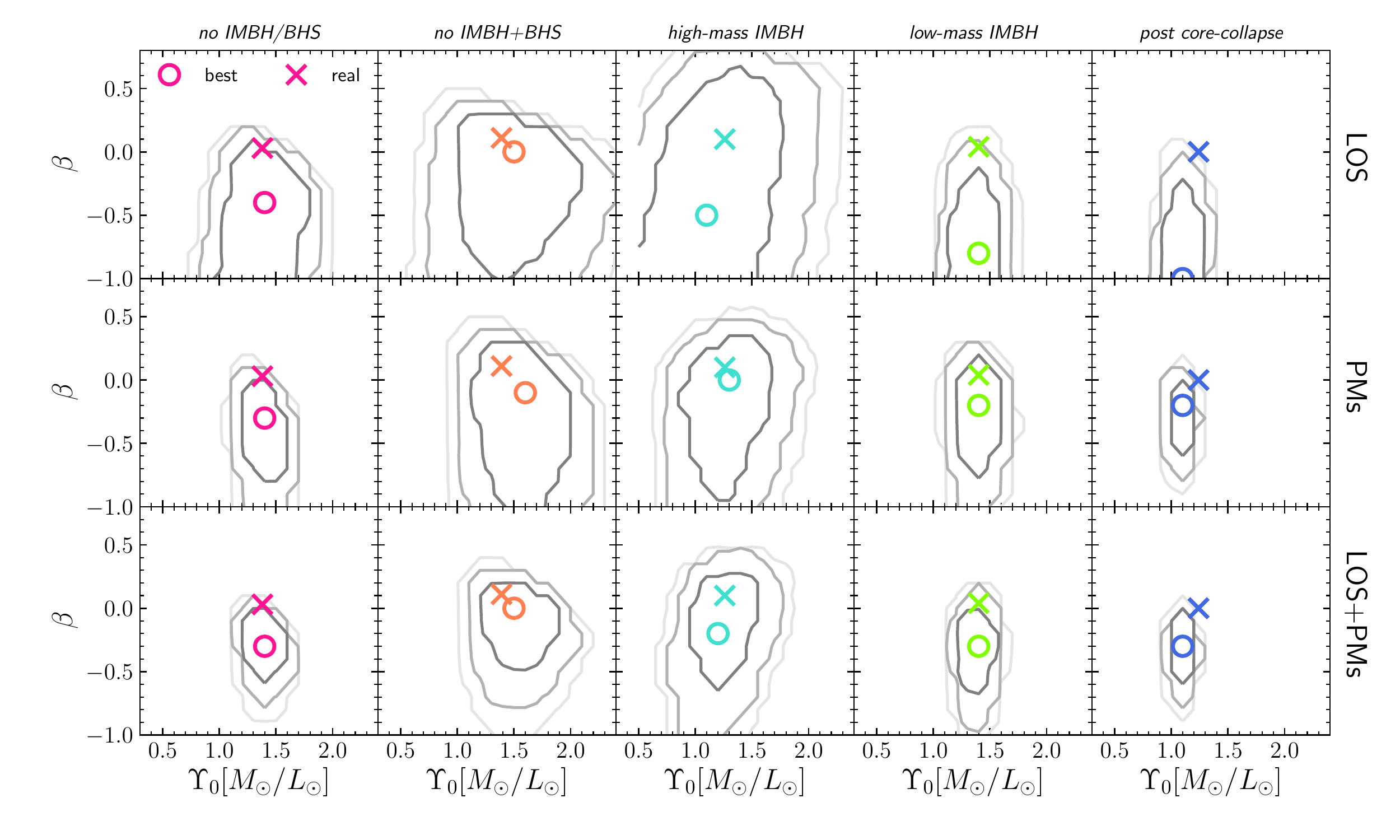}
    \caption{Parameter space for the mass-to-light ratio and velocity anisotropy, for all simulations and kinematic data used for the fit. The contours represent the confidence regions we defined to trace the errors, while the open circle  represent the best-fit value in each case and the x represent the value masured directly from the simulations within the half-mass radius. For most of the simulations the constraints improve while including more kinematic data. This is not the case for the high-mass IMBH model, where the constraints in the velocity anisotropy do not improve when including proper motions. The central shape of second velocity moment is significantly dominated by the IMBH, the changes due different velocity anisotropy values are watered-down by the presence of the high-mass IMBH.}
    \label{fig:ML_ANI_BEST}
\end{figure*}

\begin{figure*}
    \centering
    \includegraphics[width=0.8\linewidth]{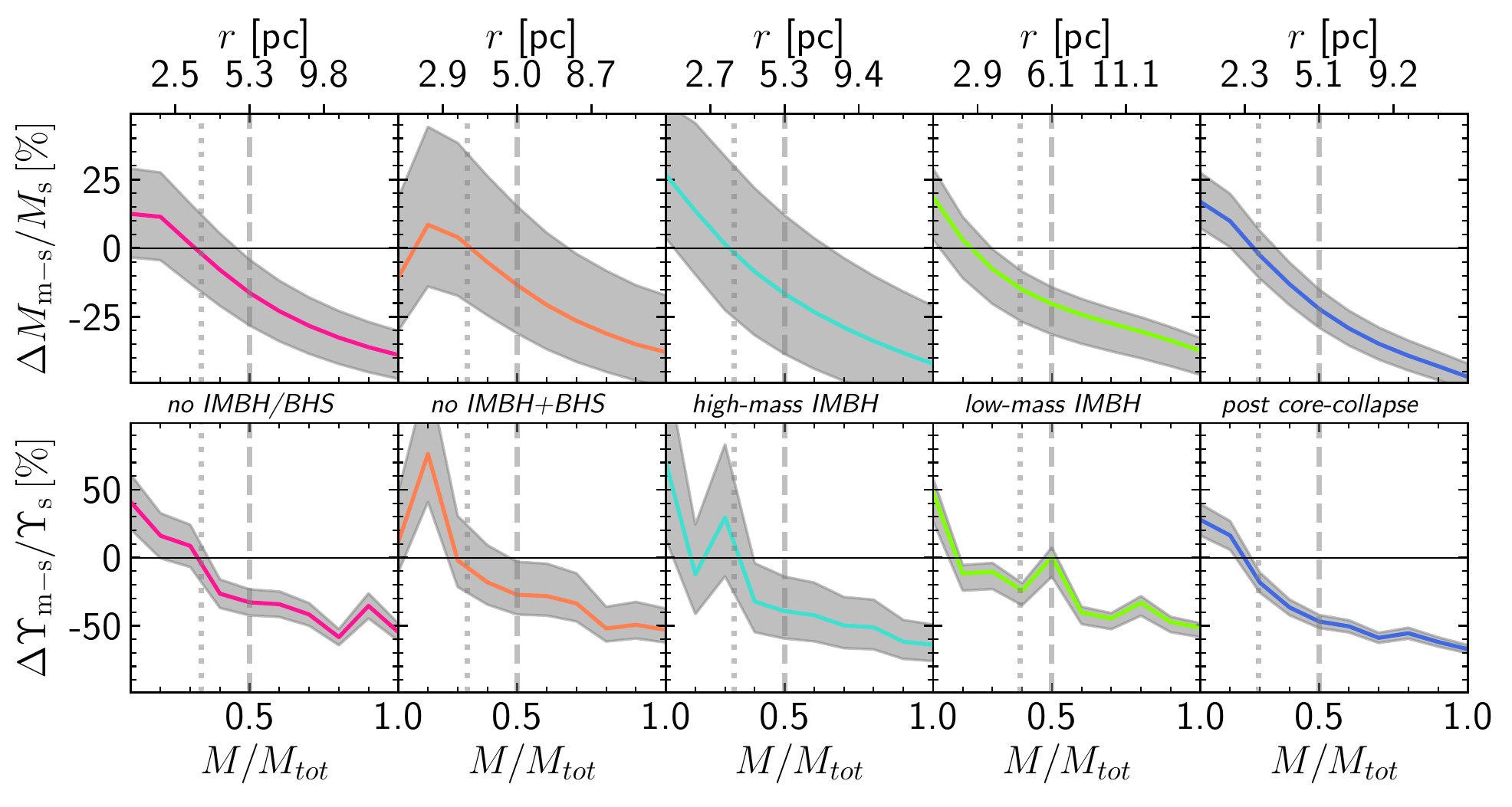}
    \caption{Mass and mass-to-light error per radius for all simulated GCs. For all plots, the x-axis is in mass-fraction of the cluster from the centre (langrangian radii). The half-mass radius is marked as a vertical dashed line, the deprojected half-light radius is marked as a dotted line. The gray area represents the range of models with $\Delta\chi^2\leq7.8$ and the coloured line represent the best fit model. On top we illustrate the values in parsec for three langrangian radii as reference. In the top panels we see that for all five simulated GCs we systematically underestimate the total mass, while overestimating the inner regions (as we represented the profiles in mass-fraction, we are unable to observe the innermost region where the IMBH is relevant). The mass profile errors behavior by radius is tightly correlated to the difference between our assumed constant mass-to-light ratio and the one from the simulation (bottom panels). In all simulated GCs, the models and the simulations are in agreement (low relative error) around the half-light radius.}
    \label{fig:mass_ml_error}
\end{figure*}

\begin{figure*}
    \centering
    \includegraphics[width=0.8\linewidth]{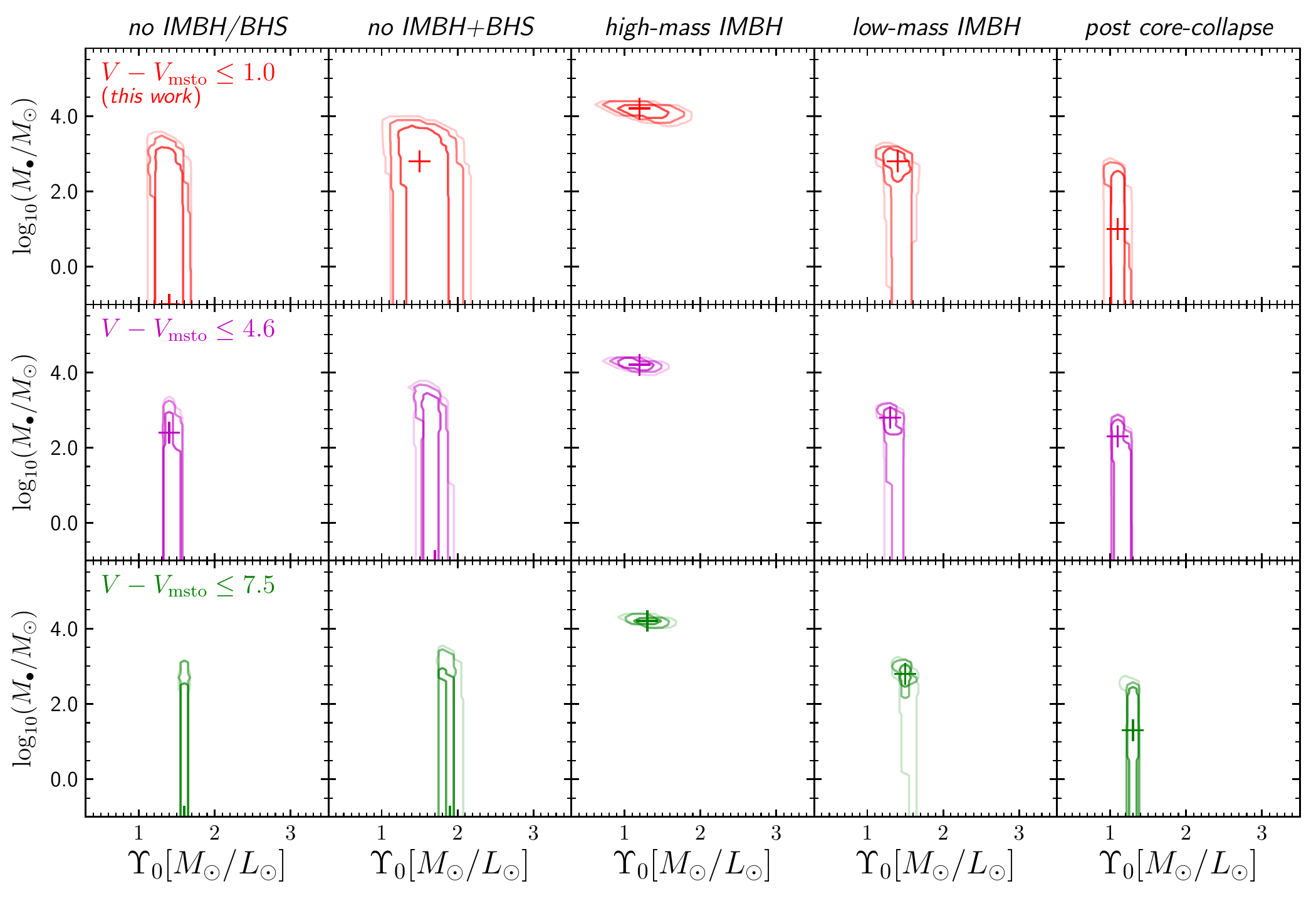}
    \caption{Constraints on the mass-to-light ratio and mass of the possible central IMBH for all simulated GCs (each column), considering the full kinematic sample (as in Figure \ref{fig:DYN-ALL-LOS+PMS}). Each row indicate a different selection sample in magnitude following the limits in Figure \ref{fig:msto_lum_cut}. The constraints are consistent for all cases. Although the second and third row are beyond the current limits for line-of-sight velocities, while the third is only possible outside $R_h$, this comparison shows that the limitations in the modelling described in this work are intrinsic to the model and do not depend on the selected sample. 
    For the \textit{high-mass IMBH} and \textit{low-mass IMBH}, the best-fit values are consistent with the expected values. On the other hand, for the three GCs without a central IMBH the best-fit values of the possible central IMBH do not converge. Once deeper observations are available allowing for a fainter limit in the luminosity cut, the Jeans modelling will automatically produce better results as our stochastic errors decrease with more stars in each bin.}
    \label{fig:fits_deeper_cuts}
\end{figure*}

\begin{figure*}
    \centering
    \includegraphics[width=0.8\linewidth]{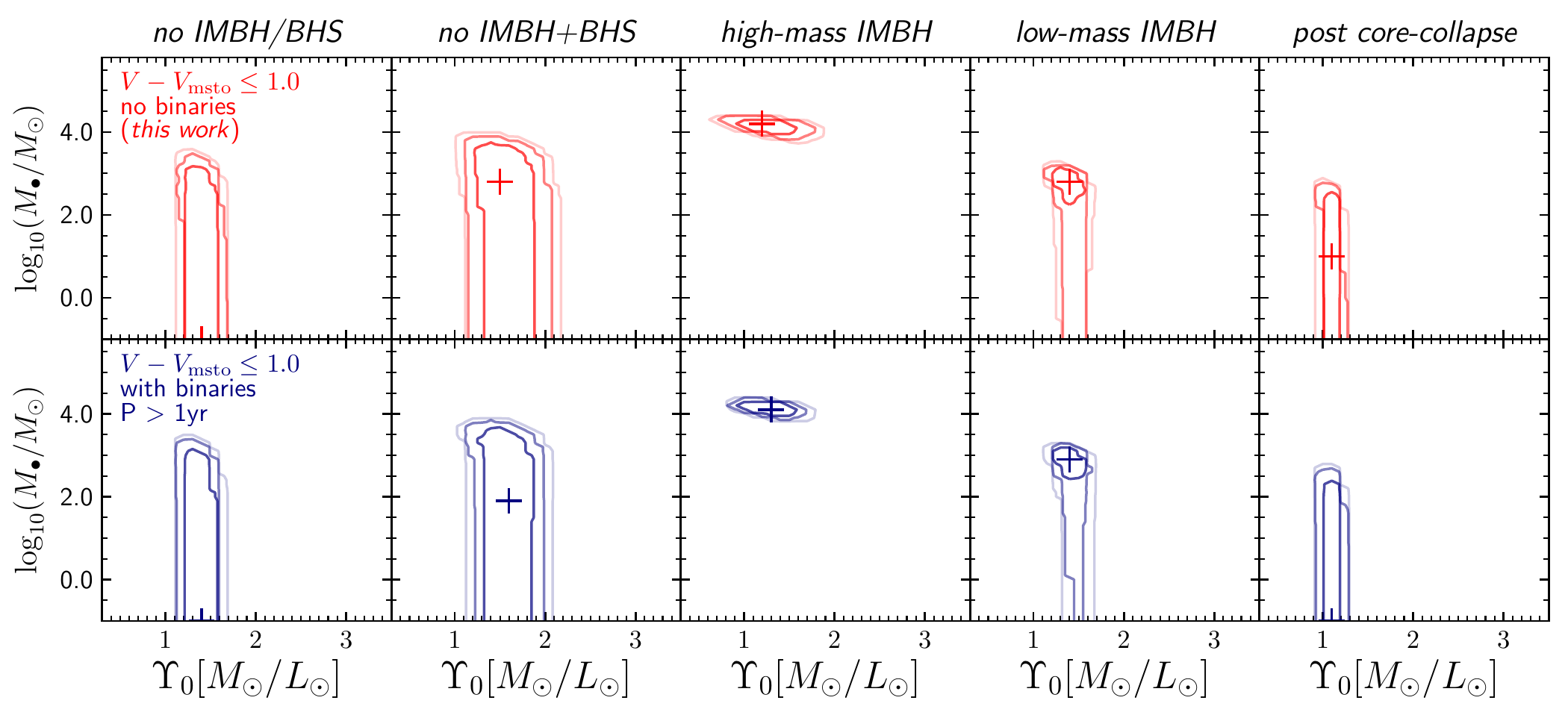}
    \caption{As in Figure \ref{fig:fits_deeper_cuts}, but considering different binary samples. The first row corresponds to the case without binaries as in our main analysis, while the bottom row shows the case when long period binaries ($P>1\,\text{yr}$) remains in the kinematic sample. The constraints from both cases are similar. As shown in panel (b) of Figure \ref{fig:BIN_CENTER}, the sample with contamination from long period binaries is consistent with the case without binaries (within errors), which is reflected on the parameter space.}
    \label{fig:fits_with_binaries}
\end{figure*}


\bsp	
\label{lastpage}
\end{document}